\documentclass[aps,superscriptaddress,eqsecnum,nofootinbib,preprintnumbers]{revtex4}
\pdfoutput=1
\usepackage{amsmath}
\usepackage{amssymb}
\usepackage{amsfonts,amsthm,bm}
\usepackage{color,xcolor}
\usepackage[greek,english]{babel}
\usepackage{graphicx,epsfig}
\graphicspath{{figures/}{./}}
\usepackage{float}
\usepackage{subfigure}
\usepackage{tikz}
\newcommand{\be}{\begin{eqnarray}}
\newcommand{\ee}{\end{eqnarray}}
\newcommand{\bea}{\begin{eqnarray}}

\newcommand{\eea}{\end{eqnarray}}

\def\PR#1{{Phys.\ Rev.\ D \bf #1}}
\def\PRL#1{{Phys.\ Rev.\ Lett.\ \bf #1}}

\begin{document}
\baselineskip=0.4 cm
\title{Spontaneous Holographic Scalarization of Black Holes in Einstein-Scalar-Gauss-Bonnet Theories}

\author{Hong Guo}
\email{gh710105@gmail.com}
\affiliation{School of Physics and Astronomy,
Shanghai Jiao Tong University, Shanghai 200240, China}
\author{Stella Kiorpelidi}
\email{stellakiorp@windowslive.com} \affiliation{Physics Division,
National Technical University of Athens, 15780 Zografou Campus,
Athens, Greece.}
\author{Xiao-Mei Kuang}
\email{xmeikuang@yzu.edu.cn}
\affiliation{Center for Gravitation and Cosmology, College of Physical Science and Technology, Yangzhou University, Yangzhou 225009, China}
\author{Eleftherios Papantonopoulos}
\email{lpapa@central.ntua.gr} \affiliation{Physics Division,
National Technical University of Athens, 15780 Zografou Campus,
Athens, Greece.}
\author{Bin Wang}
\email{wang_b@sjtu.edu.cn}
\affiliation{Center for Gravitation and Cosmology, College of Physical Science and Technology, Yangzhou University, Yangzhou 225009, China}
\affiliation{School of Aeronautics and Astronautics, Shanghai Jiao Tong University, Shanghai 200240, China}
\author{Jian-Pin Wu}
\email{jianpinwu@yzu.edu.cn}
\affiliation{Center for Gravitation and Cosmology, College of Physical Science and Technology, Yangzhou University, Yangzhou 225009, China}

\date{\today}

\begin{abstract}
We holographically investigate the scalarization in the Einstein-Scalar-Gauss-Bonnet gravity with a negative cosmological constant.  We find that instability exists for both Schwarzschild-AdS and Reissner-Nordstrom-AdS black holes with planar horizons when we have proper interactions between the scalar field and the Gauss-Bonnet curvature corrections. We relate such instability to  possible holographic  scalarization  and construct the corresponding hairy black hole solutions in the presence of the cosmological constant.  Employing the holographic principle we expect  that such bulk scalarization corresponds to the boundary description of the scalar hair condensation without breaking any symmetry, and we calculate the related holographic entanglement entropy of the system. Moreover, we compare the mechanisms of the holographic scalarizations caused by the effect of the coupling of the scalar field to the Gauss-Bonnet term  and holographic superconductor effect in the presence of an electromagnetic field, and unveil their differences in the effective mass of the scalar field, the temperature dependent property and the optical conductivity.
\end{abstract}

\maketitle
\tableofcontents

\section{Introduction}

The simplest modification of General Relativity (GR) is to introduce a scalar field  in the Einstein-Hilbert action, which is  known as  scalar-tensor theory. If the scalar field backreacts to the background metric, hairy black hole solutions are expected to be generated. There was an extensive study of the solutions and  no-hair theorems were developed constraining the possible black hole solutions and their parameters. One of the first hairy black hole solutions in an asymptotically flat spacetime was discussed in \cite{BBMB} but  it was found that there is a divergence of the scalar field on the event horizon and soon it was realized that  the solution was unstable \cite{bronnikov}. A way to avoid such irregular behaviour of the scalar field on the horizon is to introduce a scale in the gravity sector of the theory with the presence of a cosmological constant.   The resulting hairy black hole solutions  have a regular scalar field behaviour on the horizon and all the possible infinities are hidden behind the horizon \cite{Martinez:1996gn,Banados:1992wn,Martinez:2004nb,Zloshchastiev:2004ny,Martinez:2002ru,
Dotti:2007cp,Torii:2001pg,Winstanley:2002jt,Martinez:2006an,Kolyvaris:2009pc,Khodadi:2020jij}.

Without the presence of any matter hairy black hole solutions can also be generated by introducing a  coupling of a scalar field directly  to second order algebraic curvature invariants. The introduction of this coupling leads to hairy black holes because the scalar field interacts with   the spacetime curvature.  Various black hole solutions and compact objects in extended scalar-tensor-Gauss-Bonnet gravity theories were studied in the literature \cite{Mignemi_1993,Kanti_1996,Torrii_1996,Ayzenberg_2014,Kleihaus_2011,Kleihaus_2016a}. The simplest  approach is to couple the  scalar field  to the Gauss-Bonnet (GB) invariant in four dimensions.

Recently in these gravity theories choosing particular forms of the scalar coupling function, scalarized black hole solutions were generated, evading in this way  the no-hair theorems \cite{Chen:2008hk,Chen:2006ge,Doneva_2018a,Antoniou_2018,Antoniou_2018a,Doneva_2018,Silva_2018}.
In these solutions the Schwarzschild black hole background becomes unstable   below a  critical black hole mass  \cite{Doneva_2018a,Minamitsuji:2018xde,Silva_2018,Myung_2018b}, and it was found that  scalarized black holes were generated at certain masses. The characteristic feature of these solutions is that the   scalar charge is not primary, but it is connected with  the black hole mass. Various scalarized black hole solutions   and compact objects in Einstein-scalar-Gauss-Bonnet theories were discussed in \cite{Blazquez-Salcedo:2018jnn,Bakopoulos:2018nui,Silva:2018qhn,Cunha:2019dwb,Bakopoulos:2020dfg,Blazquez-Salcedo:2020rhf}.  Spontaneous scalarization of  asymptotically Anti-de Sitter  black holes in Einstein-scalar-Ricci-Gauss-Bonnet gravity  with applications to holographic phase transitions was studied in \cite{Brihaye:2019dck}.

 The scalarization of black holes in scalar-tensor-Gauss-Bonnet gravity theories when an electromagnetic field is present was studied in \cite{Doneva:2018rou}.  The presence of the GB invariant acts  as a high curvature term that triggers the instability of the background Reissner-Nordstr\"{o}m black hole. Then the coupling of the scalar field to these terms scalarized the Reissner-Nordstr\"{o}m black hole.  In spite that the electromagnetic field is not coupled to the GB term, the presence of the charge $Q$   is introducing  another scale in the theory, and the interplay between the mass $M$ and the charge $Q$ gives more  interesting results on the scalarization behaviour of the  Reissner-Nordstr\"{o}m black hole  compared to that of the Schwarzschild case.

In this work we will study the spontaneous scalarization of planar  black holes when  there is a coupling of a scalar field to the GB term in a scalar-tensor theory with a negative cosmological constant and its holographic realization.
 In section \ref{secII}, we will show that the presence of the coupling of the scalar field to the curvature invariant introduces an extra negative contribution to the effective mass of the scalar field, and then the scalar potential emerges the potential well, leading to an instability. We will analyze the instability of both Schwarzschild-AdS (SAdS) and Reissner-Nordstrom-AdS black holes with planar horizon. For the case with neutral scalar, comparing to that in Schwarzschild-flat black holes, the potential well modified by the GB coupling in our AdS case is larger and deeper, which implies that the scalarization can occur more easily.

We further investigate the bulk formation process of a hairy black hole due to the  GB coupling in the case with neutral scalar field. Moreover, according to the remarkable gauge/gravity duality \cite{Maldacena:1997re,Gubser:1998bc,Witten:1998qj}, a black hole in the gravity side in AdS spacetime is holographically dual to a certain state in the dual conformal field theory (CFT), thus, the SAdS black hole and the hairy AdS black hole should be dual to different states in the boundary CFT theory. Consequently, from the CFT side, the spontaneous scalarization in the bulk can be interpreted as a certain phase transition, and the GB coupling somehow mimics a  mechanism that leads to the phase transition. In addition, according to holography, if we perturb a scalar field and we find unstable modes then these modes correspond to an expectation value for ${\cal O}_\phi$, and such an expectation value will condensate. To further understand the scalarized process and its holographic duality, we study how the condensation of ${\cal O}_\phi$ emerges as the GB coupling increases  from CFT side and then probe the dual phase transition via holographic entanglement entropy.
 The occurrence of phase transition is usually accompanied by symmetry breaking, unless it is a quantum phase transition. The obtained phase transition dual to the spontaneous scalarization of SAdS black hole due to the GB coupling is likely to be a certain quantum phase transition since there is no related symmetry in the setup.

It is well known that a hairy black hole can be formed  below a critical temperature because of the condensation of a charged scalar field coupling to a Maxwell field. Such mechanism  was described by Gubser in
\cite{Gubser:2005ih,Gubser:2008px} where it was shown that a spontaneous breaking of the $U(1)$ symmetry  leads to  the occurrence of a holographic superconducting phase transition \cite{Hartnoll:2008vx,Hartnoll:2008kx}.  However, it is clear that the formation of the hairy black hole because of interaction between the scalar field and GB curvature correction is different from such Abelian Higgs model. We shall discuss the differences both in the bulk gravity side and on the boundary CFT side. The details will be present in section \ref{sectIII}.

In addition to discussing differences in two mechanisms of scalarizations, in section \ref{sectIV}, we will add the $U(1)$ gauge field to the Einstein-Scalar-Gauss-Bonnet theory, such that the scalar field becomes charged and we will investigate the  charged scalar field condensation in the background of Schwarzschild-AdS planar black hole. This can present us a picture on the combined effect of two different scalarization mechanisms, which can accommodate a wider and deeper effective mass to speed up the formation of hairy black holes. We will observe that above certain critical temperature, only the GB coupling plays the role in the formation of scalar hair, however, when temperature drops below critical value, the holographic superconducting condensation participates and we have the combined stronger effects on the formation of hairy black holes. Finally, we present our conclusion and discussion in section \ref{sectCon}.

%The work is organized as follows. In Section \ref{sectII} we study the stability of a  Gauss Bonnet theory coupled to a scalar field in four dimensions.
%In Section  \ref{sectIII} we discuss in the probe limit the scalarization in the gravity sector with a $U(1)$ gauge field, the formation of a condensation and we calculate the optical conductivity. In Section \ref{sectIV} we study the scalarization with a neutral scalar field both in probe limit and with the scalar field backreacting to the background Schwarzschild-AdS black hole. We also study the entanglement entropy as a probe to the scalarization.  In Section \ref{sectVI} are our conclusions.

\section{Stability analysis of a  Gauss-Bonnet theory coupled to a scalar field  }\label{secII}
\subsection{Model and setup}

 In this section we will study the stability of a  Gauss-Bonnet theory coupled to a scalar field  in 4-dimensions. Let us first consider a
charged  scalar field $\phi$ outside   the horizon of a black hole described by the Lagrangian
\be
  {\cal L} =\frac{1}{16\pi G_N }\Big{(} R + {6 \over L^2} - {1 \over 4} F_{\mu\nu}^2 -
    |\nabla_\mu \phi - i q A_\mu \phi|^2 - m^2 |\phi|^2 \Big{)}\,, \label{lagr}
\ee
where $G_N$ is the   Newton's constant, $F_{\mu\nu}$ is the Maxwell field, $L$ is the curvature radius of AdS spacetime and $\phi$ is a real scalar field with mass $m$ and charge $q$.

Consider a charged black hole as the  background metric. Then, the metric, the electromagnetic field and the scalar field are
\be
  ds^2 = g_{tt} dt^2 + g_{rr} dr^2 + ds_2^2 \nonumber \\
  A_\mu dx^\mu = \Phi(r) dt,  \, \, \, \, \phi = \phi(r) \,, \label{11}
 \ee
 where all fields are assumed to depend only on $r$. Assuming that there is no backreaction of the scalar field to the metric the scalar field part of the Lagrangian (\ref{lagr}) is
  \be
  {\cal L}_\phi = \frac{1}{16\pi G_N }\Big{(}-g^{tt} q^2 \Phi^2 |\phi|^2 -
    g^{rr} |\partial_r \phi|^2 -
    m^2 |\phi|^2 \Big{)} \,.
 \ee
 Then the effective mass of the scalar field $\phi$ is
 \be
  m_{\rm eff}^2 = m^2 + g^{tt} q^2 \Phi^2 \,. \label{effmas}
\ee
Because $g^{tt}$ is negative outside the horizon then  $m_{\rm eff}^2$ should become negative there if $\Phi$ is non-zero. If $q$ is large, and the electric field outside the horizon is large and $m^2$ is small, then  $m_{\rm eff}^2$  becomes negative a little outside the horizon. Then it was argued in \cite{Gubser:2008px,Gubser:2005ih} this is a signal of an instability. However, if this instability could be manifest at large distances and not only outside the horizon then  one has to study the superradiance effect \cite{Penrose:1971uk,Bekenstein} and rely on the qausinormal modes (QNMs) of the perturbed scalar field in the background of the charged black hole.

Let us now consider the scalar-tensor theory with the scalar field coupled to GB term. Because of the presence of the coupling of the scalar field to the GB term, eq. (\ref{effmas}) will be modified and this coupling will appear as another source of instability with important consequences in the theory as we will discuss in the following.

 Consider a theory which is given by the action \cite{Doneva:2018rou}
\begin{equation}
    S=\frac{1}{16 \pi G_N} \int \, d^4 x \, \sqrt{-g} \left[R+\frac{6}{L^2}-\frac{1}{4}F_{\mu\nu}F^{\mu \nu}-D_\mu \phi (D^\mu \phi)^*-m^2 |\phi|^2 +f(\phi) \mathcal{R}^2_{GB}\right]~,\label{action}
\end{equation}
where $D_\mu=\nabla_\mu-iqA_\mu $, $\phi=\phi(r)$ is set to be real scalar field and $f(\phi)$ is the scalar field coupling function which depends only on $\phi$ and the GB term is given by
\begin{equation}
    \mathcal{R}^2_{GB}=R^2-4R_{\mu\nu}R^{\mu\nu}+R_{\mu\nu\alpha\beta}R^{\mu\nu\alpha\beta}~.
\end{equation}
The variation of the action $(\ref{action})$ with respect to the metric yields the modified Einstein equations as
\begin{align}
R_{\mu \nu}-\frac{1}{2}R g_{\mu\nu}-\frac{3}{L^2}g_{\mu\nu}+\Gamma_{\mu\nu}=&\nabla_\mu \phi\nabla_\nu\phi-\frac{1}{2}g_{\mu\nu}\nabla_\alpha \phi \nabla^\alpha\phi-\left(\frac{1}{2}m^2g_{\mu\nu}+\frac{1}{2}q^2A_\alpha A^\alpha g_{\mu\nu}-q^2A_\mu A_\nu\right)\phi^2\nonumber\\
&+\frac{1}{2}\left(F^{\mu \alpha}F^\nu{}_\alpha-\frac{1}{4}F_{\alpha \kappa}F^{\alpha\kappa}g_{\mu\nu}\right)~, \label{einsteinequation}
\end{align}
where $\Gamma_{\mu\nu}$ is defined by
\begin{align}
\Gamma_{\mu\nu}=&-R\left(\nabla_\mu \Psi_\nu+\nabla_\nu\Psi_\mu\right)-4\nabla^\alpha\Psi_\alpha\left(R_{\mu\nu}-\frac{1}{2}Rg_{\mu\nu}\right)+4R_{\mu\alpha}\nabla^\alpha\Psi_\nu+4R_{\nu\alpha}\nabla^\alpha\Psi_\mu\nonumber\\
&-4g_{\mu\nu}R^{\alpha\beta}\nabla_\alpha\Psi_\beta+4R^\beta{}_{\mu\alpha\nu}\nabla^\alpha\Psi_\beta~,
\end{align}
with
\begin{equation}
\Psi_\mu=\frac{df(\phi)}{d\phi} \nabla_\mu\phi~.
\end{equation}
The variation of the metric $(\ref{action})$ with respect to the electric potential leads to
\begin{equation}
\nabla_\alpha F^{\alpha\mu}=2q^2A^\mu \phi^2~,
\end{equation}
while the Euler-Lagrange equation for the scalar field and Maxwell electromagnetic field reads as
\begin{eqnarray}
\nabla_\mu \nabla^\mu \phi -\left(m^2+q^2A_\mu A^\mu \right)\phi+\frac{1}{2}f'(\phi) \mathcal{R}^2_{GB}=0~. \label{scalarfieldequation}
\end{eqnarray}

Note that different choices of the function $f (\phi)$ correspond to different scalar GB gravity theories \cite{Doneva_2018a}. A function like $f (\phi) \sim 1 - e^{-\phi^2}$ satisfies the conditions for spontaneous scalarization for a trivial scalar field, namely $f'(\phi_0)=0$ and $f''(\phi_0)>0$  \cite{Doneva_2018a}. So here we choose
\begin{equation}
    f (\phi) =\frac{\lambda^2}{2\beta}\left( 1- e^{-\beta\phi^2}\right)~.\label{function1}
\end{equation}
Note that for convenience we have incorporated  the Gauss-Bonnet coupling $ \lambda $ in the $ f (\phi) $ function (\ref{function1}). Moreover, we found that our results are slightly affected by the parameter $\beta$, so we will set $\beta=1$ in the following study.
If we consider that the scalar field and the Maxwell field do not backreact to the metric then the Einstein equations admit the planar Schwarzschild-AdS black hole as a solution
\begin{equation}\label{background}
ds^2=-g(r) dt^2+\frac{1}{g(r)}dr^2+r^2\left(dx^2+dy^2\right)~,
\end{equation}
where the metric function is
\begin{equation}\label{backgroundA}
g_S(r)=\frac{r^2}{L^2}-\frac{M}{r}~.
\end{equation}
In the case where  the scalar field does not backreact to the metric we get the  Reissner-Nordstr\'om-AdS black hole as a solution endowed with an electric potential
\begin{equation}
\Phi(r)=\frac{Q}{r_H} -\frac{Q}{r}~, \label{electricpotential}
\end{equation}
where the  metric function now is
\begin{equation}
    g_{RN}(r)=\frac{r^2}{L^2}-\frac{2M}{r}+\frac{Q^2}{4r^2}.\label{RNAdSmetric}
\end{equation}
Thus, the Gauss-Bonnet curvature can be calculated with the planar Schwarzchild-AdS metric $g_S$ or the Reissner-Nordstr\'om-AdS metric $g_{RN}$ as
\begin{equation}\label{geometries}
  \mathcal{R}^2_{GB}=\frac{4}{r^2}\left[g'(r)^2+g(r)g''(r)\right].
\end{equation}

\subsection{(In)stability  analysis for a charged scalar field }\label{sectIII}

A small scalar fluctuation on the background is governed by the Klein-Gordon equation
\begin{equation}
  \left( \Box  - \left(m^2+q^2A_\mu A^\mu\right)+\frac{1}{2} f''(\phi_0) \mathcal{R}^2_{GB}\right)\delta\phi=0~,\label{kleingordon}
\end{equation}
where $\Box $ is the d'Alambert operator.
This equation defines an effective mass term given by
\begin{equation}
    m^2_{eff}=m^2-\frac{q^2 \Phi(r)^2}{g(r)}-\frac{\lambda^2}{2} \mathcal{R}^2_{GB}~. \label{meffcombact}
\end{equation}
These small scalar fluctuations may break the $U(1)$ symmetry if the $m^2_{eff}$ is negative enough for long enough. This mass becomes tachyonic and the fluctuations are unstable. So it is the effective mass that determines the qualitative and quantitative behaviour of the modes of the scalar fluctuations. The mass of the black hole can be expressed as $M_S =r_h^3/2 L^2 $ for the neutral background or can be expressed as $ M_{RN}=(L^2Q^2+4L^2r_h^2+4r_h^4)/8L^2r_h~,$ for the charged one. Note that the parameter $r_h$ corresponds to the horizon of the black hole.  Moreover in both cases by using $(\ref{electricpotential})$ for the electric potential, the effective mass $(\ref{meffcombact})$  can be expressed as a function of six parameters $m_{eff}^2=m_{eff}^2(m,q,L,Q,\lambda,r)$, where we have set $r_h=1$ in both cases which guarantees that $r=1$ is the largest root of the metric function.  Moreover, in the following discussion  we shall set $L=1$ unless we assign it.

\begin{figure}[h]
    \centering
    %\subfloat[The Reissner-Nordstrom AdS background.]
    {{\includegraphics[width=8cm]{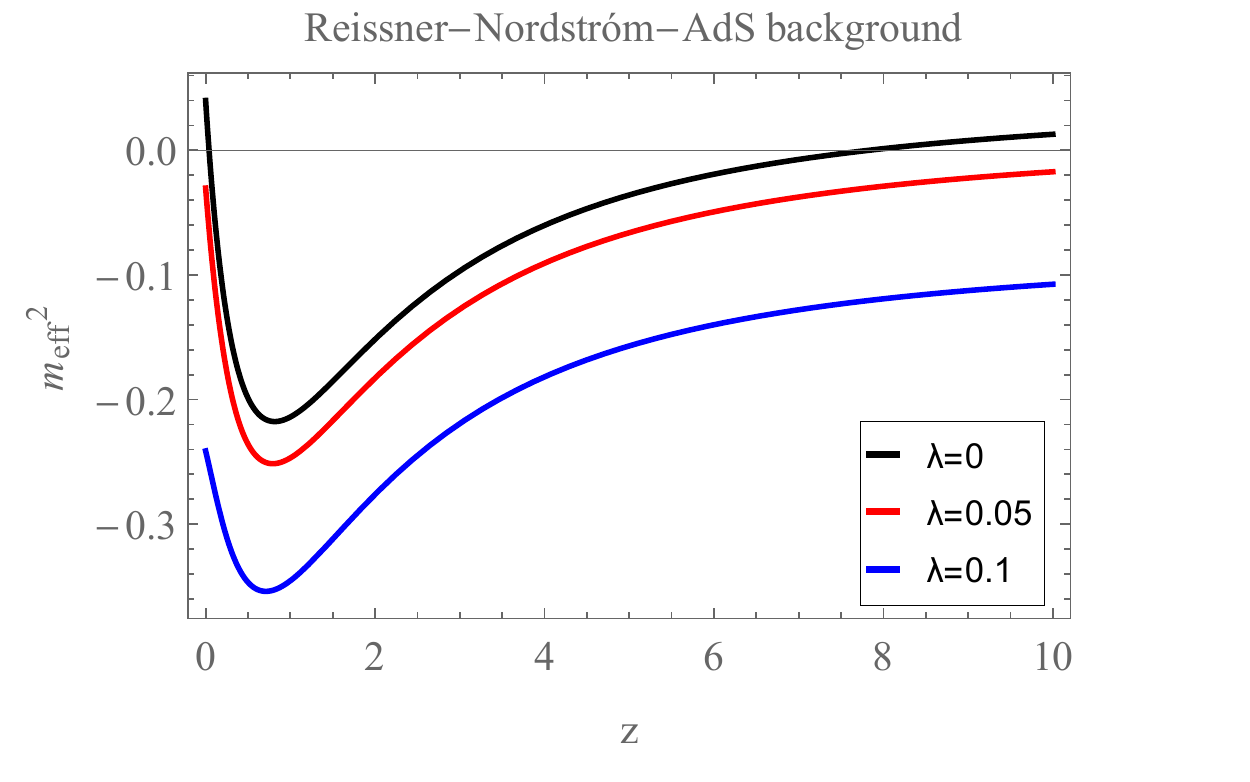} }}%
    \qquad
   % \subfloat[The Schwarzchild AdS background.]
    {{\includegraphics[width=8cm]{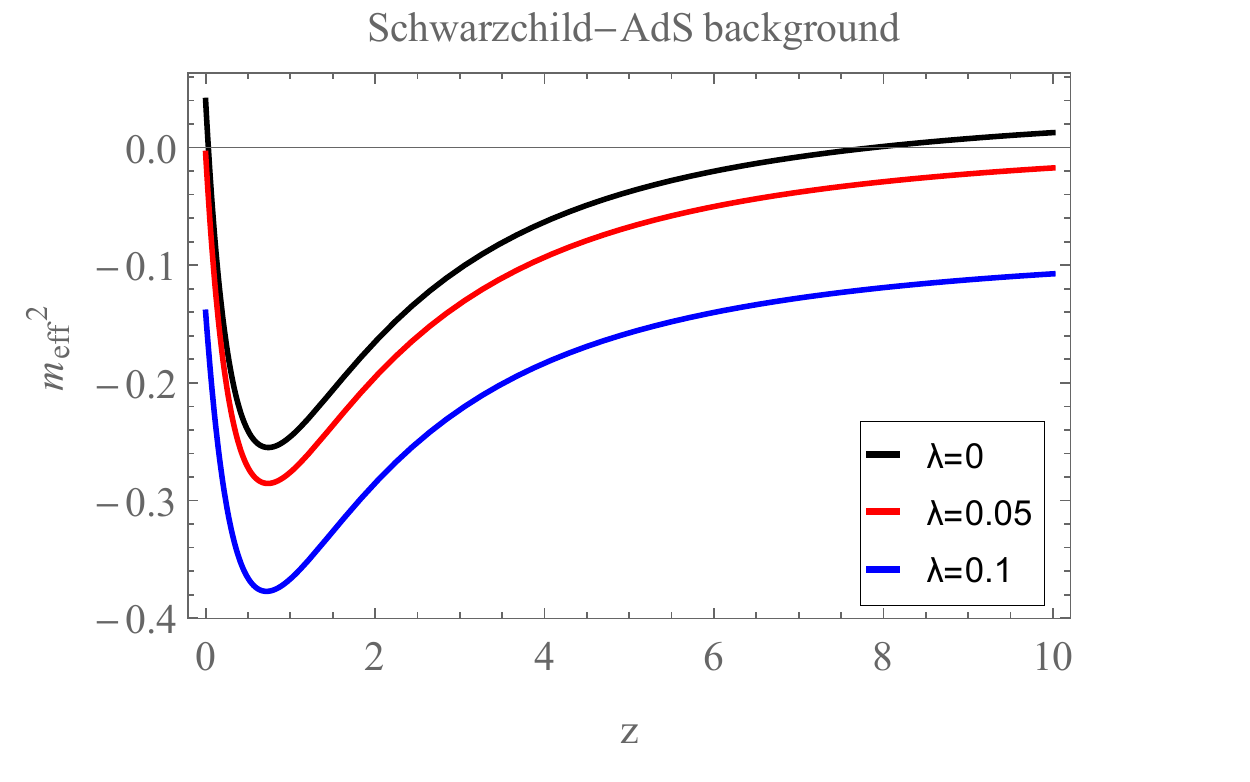} }}%
    \caption{Left: the effective mass $m^2_{eff}$ as a function of the coordinate $z=r-1$ outside the horizon (located at $z = 0$) for coupling constant $\lambda = 0, 0.05, 0.1$, the event horizon is fixed as $r_H=1$  and the other parameters are fixed as $m=0.2,q=2,L=1,Q=1$. Right: the effective mass $m^2_{eff}$ for the Schwarzschild-AdS black hole for the same value of parameters. } %
    \label{meff_lamda}
\end{figure}

\begin{figure}[h]
    \centering
    %\subfloat[The Reissner-Nordstrom AdS background.]
    {{\includegraphics[width=8cm]{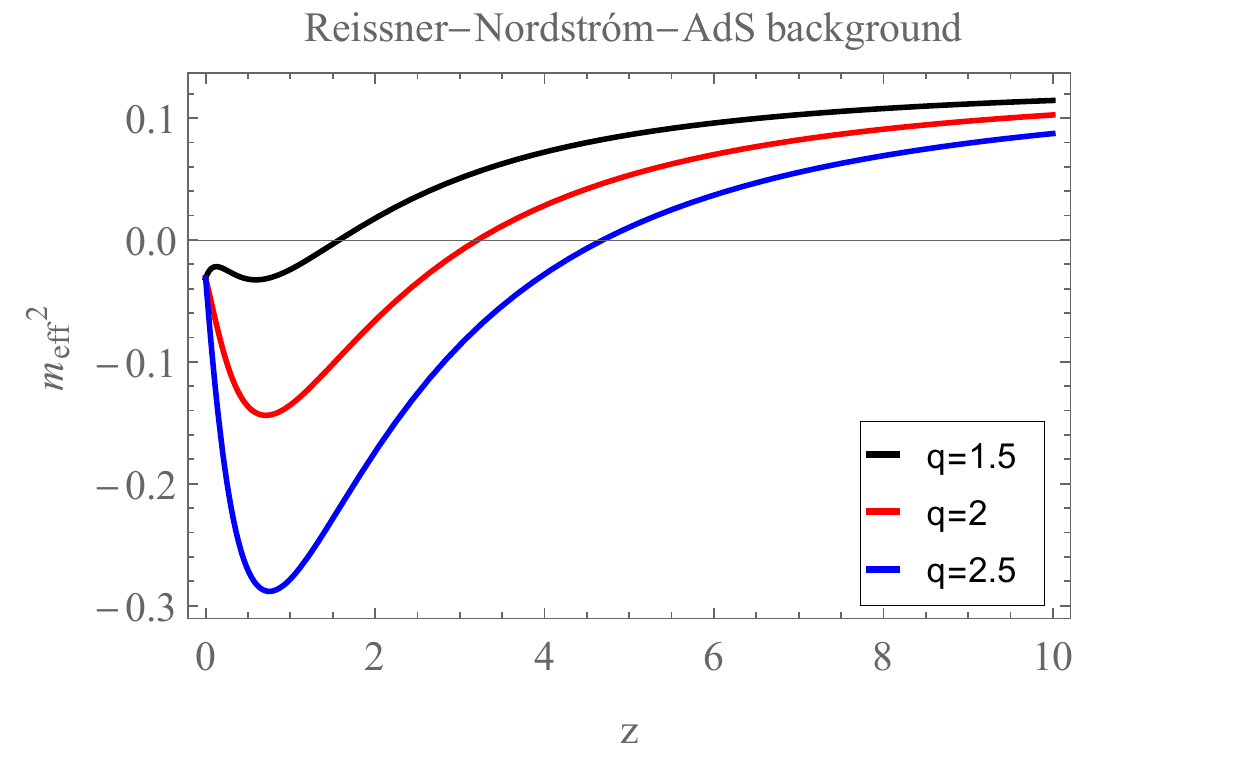} }}%
    \qquad
    %\subfloat[The Schwarzchild AdS background.]
    {{\includegraphics[width=8cm]{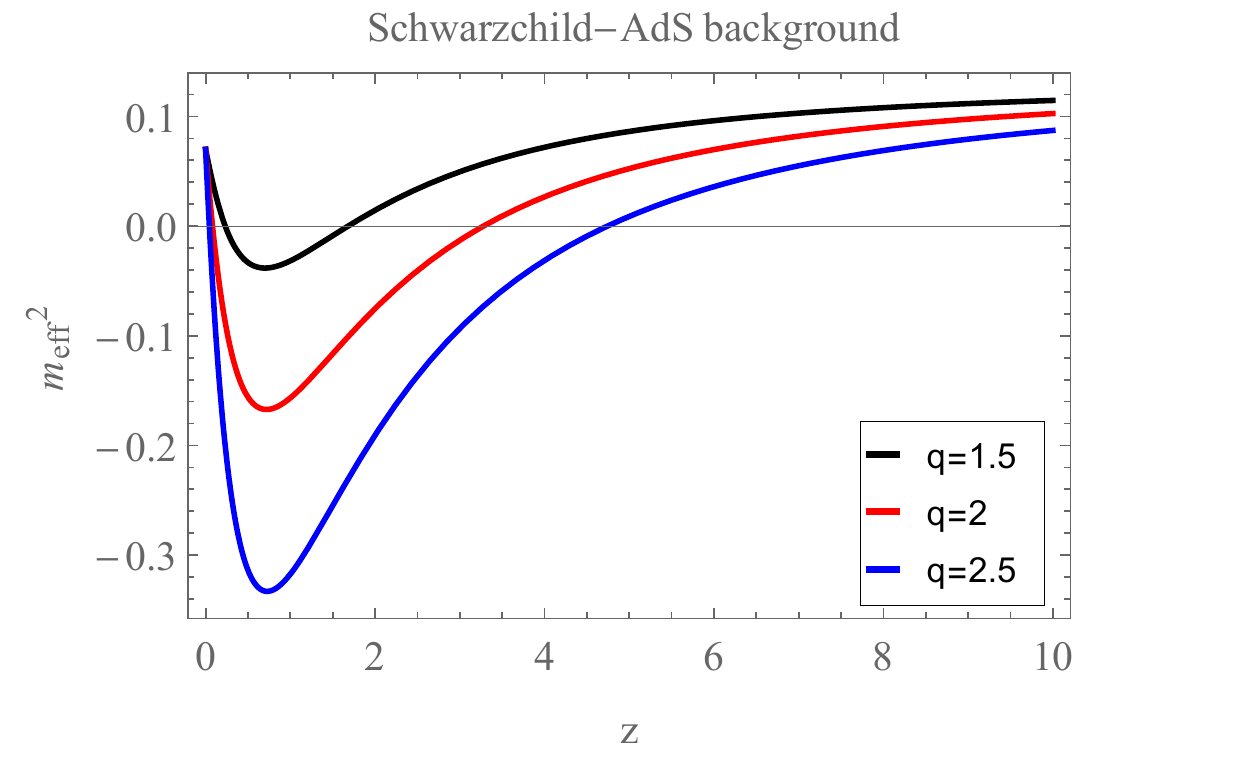} }}%
    \caption{Left: the effective mass $m^2_{eff}$ as a function of the coordinate $z=r-1$ outside the horizon (located at $z = 0$) for scalar charge $q = 1.5, 2, 2.5$, the event horizon is fixed as $r_h=1$  and the other parameters are fixed as $m=0.5, \lambda=0.1,L=1,Q=1$. Right: the effective mass $m^2_{eff}$ for the Schwarzschild-AdS black hole for the same value of parameters. }
      \label{meffq}
\end{figure}

\begin{figure}[h]
    \centering
   % \subfloat[The Reissner-Nordstrom AdS background.]
    {{\includegraphics[width=8cm]{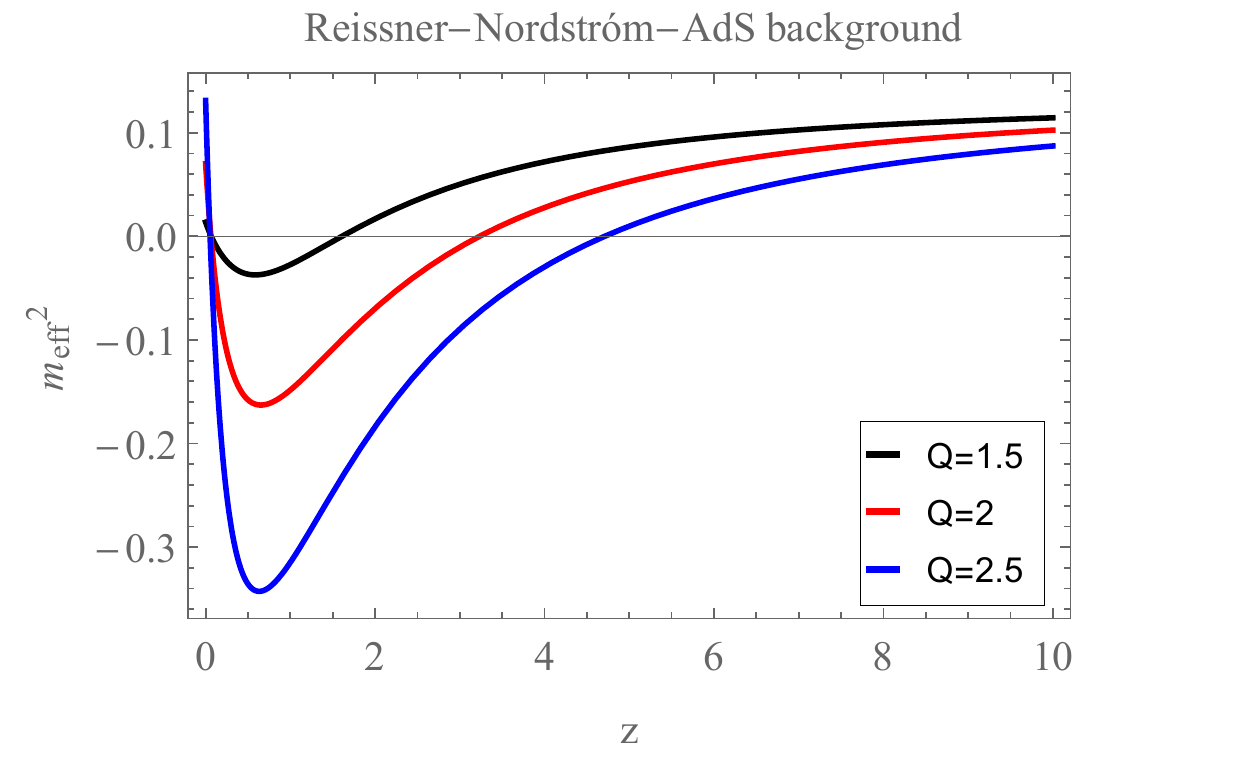} }}%
    \qquad
    %\subfloat[The Schwarzchild AdS background.]
    {{\includegraphics[width=8cm]{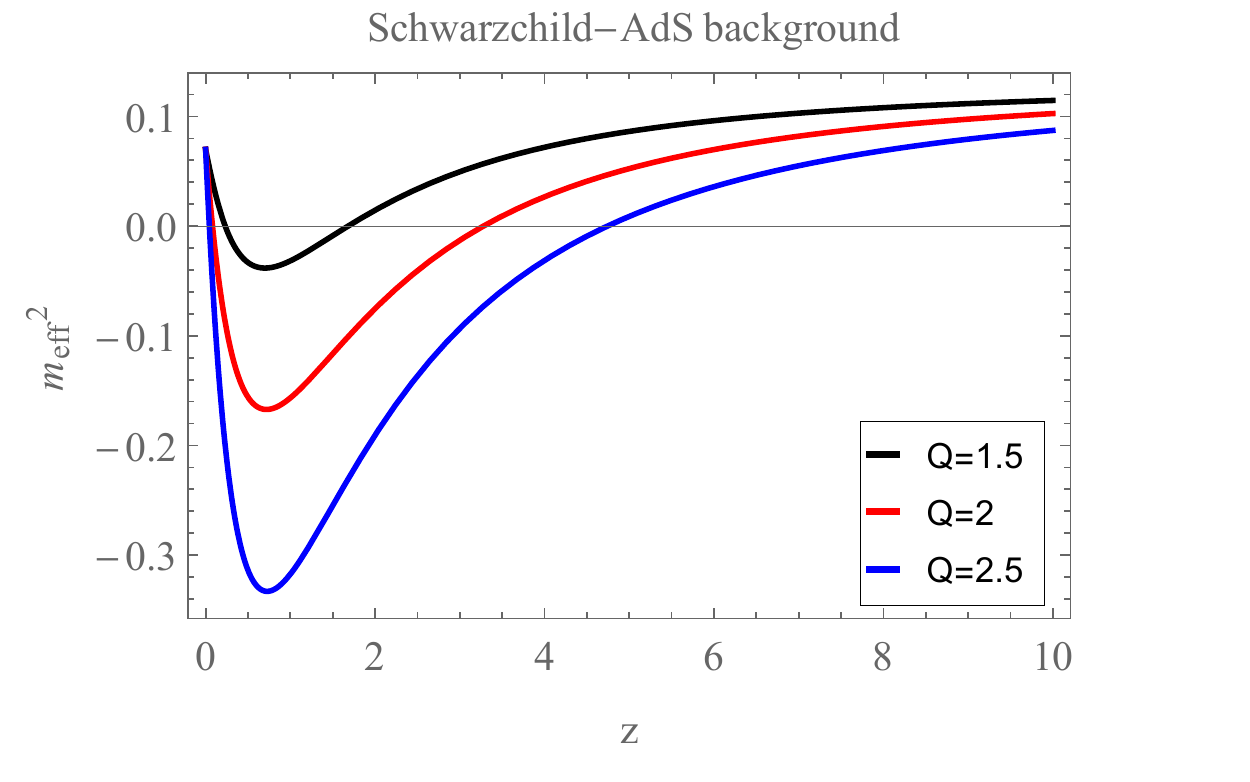} }}%
    \caption{Left: the effective mass $m^2_{eff}$ as a function of the coordinate $z=r-1$ outside the horizon (located at $z = 0$) for black hole charge $Q = 1.5, 2, 2.5$, the event horizon is fixed as $r_h=1$  and the other parameters are fixed as $m = 0.5, q = 1,L=1,\lambda=0.1$. Right: the effective mass $m^2_{eff}$ for the Schwarzschild-AdS black hole for the same value of parameters. }%
    \label{meffQQ}%
\end{figure}

  As we can see in Fig. \ref{meff_lamda} if the GB coupling $\lambda$ is increased the effective mass becomes more negative. We observe the same behaviour if the background metric is Schwarzschild-AdS. The same behaviour for the effective mass we observe in Fig. \ref{meffq} for the charge of the scalar field $q$ and for the charge of the black hole in Fig. \ref{meffQQ}. Then  as we discussed in the previous Section, according to the gauge/gravity duality, it is more easy for the scalar field to condense.

%\begin{figure}[h]
 %   \centering
  %  \includegraphics[scale=0.7]{meff_lamda.pdf}
   % \caption{The effective mass $m^2_{eff}$ as a function of the coordinate $z=r-1$ outside the horizon (located at $z = 0$) for coupling constant $\lambda = 0, 0.001, 0.01, 0.05, 0.1$, the event horizon is fixed as $r_H=1$  and the other parameters are fixed as $m=0.2,q=2,L=1,Q=1$. }
    %\label{meff_lamda}
%\end{figure}

%\begin{figure}[h]
  %  \centering
   % \includegraphics[scale=0.7]{meff_q.pdf}
    %\caption{The effective mass $m^2_{eff}$ as a function of the coordinate $z=r-1$ outside the horizon (located at $z = 0$) for scalar charge $q = 1.5, 2, 2.5$, the event horizon is fixed as $r_H=1$  and the other parameters are fixed as $m=0.5, \lambda=0.1,L=1,Q=1$.}
    %\label{meff_q}
%\end{figure}

%\begin{figure}[h]
 %   \centering
  %  \includegraphics[scale=0.7]{meff_QQ.pdf}
   % \caption{The effective mass $m^2_{eff}$ as a function of the coordinate $z=r-1$ outside the horizon (located at $z = 0$) for black hole charge $Q = 1.5, 2, 2.5$, the event horizon is fixed as $r_H=1$  and the other parameters are fixed as $m = 0.5, q = 1,L=1,\lambda=0.1$.}
    %\label{meff_QQ}
%\end{figure}

According to the gauge/gravity duality, for a condensate to be formed on the boundary theory,  the effective mass should be  below the  Breitenlohner-Freedman (BF) bound, i.e. for  $m^2_{eff}<m^2_{BL}$.
  %the mode is effectively space-like and produces a tachyonic instability.
  In our case the $(3+1)$-dimensional asymptotically AdS spacetime with AdS radius $L=-\frac{3}{\Lambda}$ the BF bound is $m^2_{BF}=-\frac{d^2}{4 L^2}=-\frac{9}{4L^2}$. So the condition for instability is
\begin{equation}
    m^2_{eff}<-\frac{9}{4 L^2}~,
\end{equation}
\begin{equation}
    m^2-\frac{q^2 \Phi(r)}{g(r)}-\frac{\lambda^2}{2} \mathcal{R}^2_{GB}<-\frac{9}{4 L^2}~.
\end{equation}

The scalar perturbation corresponds to a particular perturbation of a thermal state of the CFT. The Hawking temperature of the background is
\begin{equation}
T=\frac{g'(r_H)}{4 \pi}~.
\end{equation}

 In the next  Section we will try to understand where this instability leads to. As we already discussed, in \cite{Gubser:2008px} it was claimed that the spontaneous breaking of an $U(1)$ symmetry  could lead to a superconducting layer formed outside the horizon of a charged black hole. On the Gravity sector this can be understood if the scalar field backreacts with the background metric and spontaneously scalarizing it. This can happens because the gravitational attraction is balanced by the electromagnetic repulsion of the scalar field bounced off the AdS boundary and then a thin layer of matter is formed outside the horizon of the black hole. In the field theory on the boundary the scalarization of the background metric corresponds to the formation of a condensate which breaks the gauge invariance of a gauge field, scalarizing it in a sense.

\subsection{(In)stability  analysis for a neutral  scalar field }

The presence of the coupling of the scalar field to the GB term introduces another term in the effective mass of the scalar field in (\ref{meffcombact}).
Then it would be interesting to study possible instabilities in the case that the scalar field is neutral and the background metric is the Schwarzschild Anti de-Sitter black hole. The GB term  is a high curvature term, so it would be interesting to see if the gravitational attraction works as  counter-balance mechanism leading to the scalarization of the bulk black hole  and the formation of a condensate on the boundary.

Consider a Schwarzschild Anti de-Sitter (SAdS) black hole
\begin{equation}
ds^2=-g(r)dt^2+\frac{dr^2}{g(r)}+r^2(dx^2+dy^2)~,\label{eq-BG}
\end{equation}
with $g(r)=-\frac{M}{r}+\frac{r^2}{L^2}$.
To study the (in)stability of the scalar field near the event horizon we have to  analyze the  effective potential and the time evolution. To this end, we consider the time-dependent radial perturbation in the background of the metric \eqref{eq-BG} as $\varphi=\epsilon \phi(r,t)/r$, and then the Klein-Gordon equation under the tortoise coordinate $r_*=\int g^{-1}dr$ takes the form
\begin{equation}
	-\frac{\partial^2 \phi(r,t)}{\partial t^2}+\frac{\partial^2 \phi(r,t)}{\partial r_{*}^2}-V(r) \phi(r,t)=0~,
\end{equation}
where the effective potential is given by
\begin{equation}
	V(r)=g\left[\frac{g'}{r}+m^2-\frac{2\lambda^2}{r^2}(g'^2+gg'')\right]~,
\end{equation}
from which the scalar mass together with the Gauss-Bonnet term contributes to the effective mass, $m_e$, of the perturbed scalar field , which we will formally define in Eq.(\ref{boundary}).

%%%%%
\begin{figure}[h]
\center{
  \includegraphics[scale=0.38]{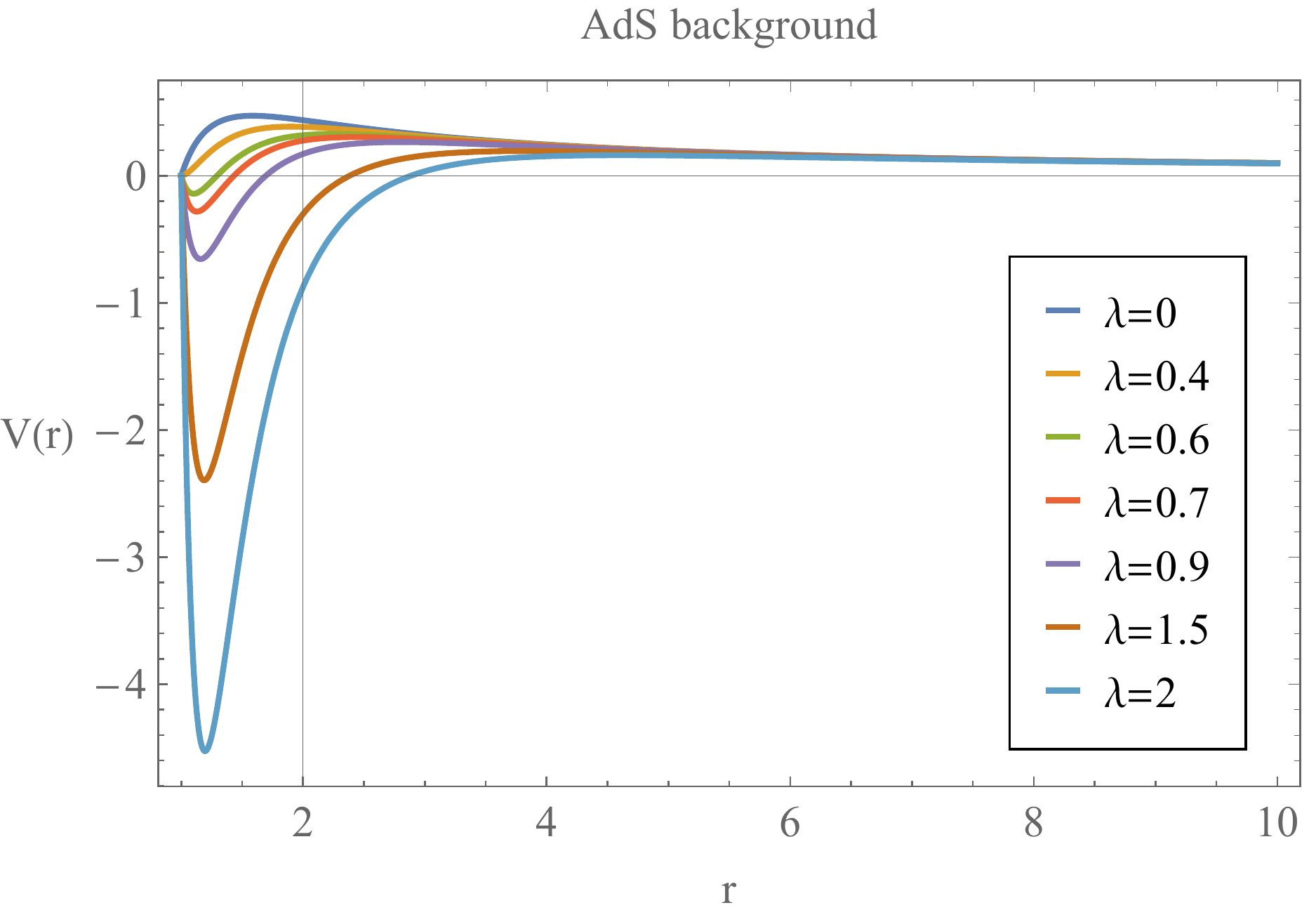}\hspace{0.5cm}
  \includegraphics[scale=0.4]{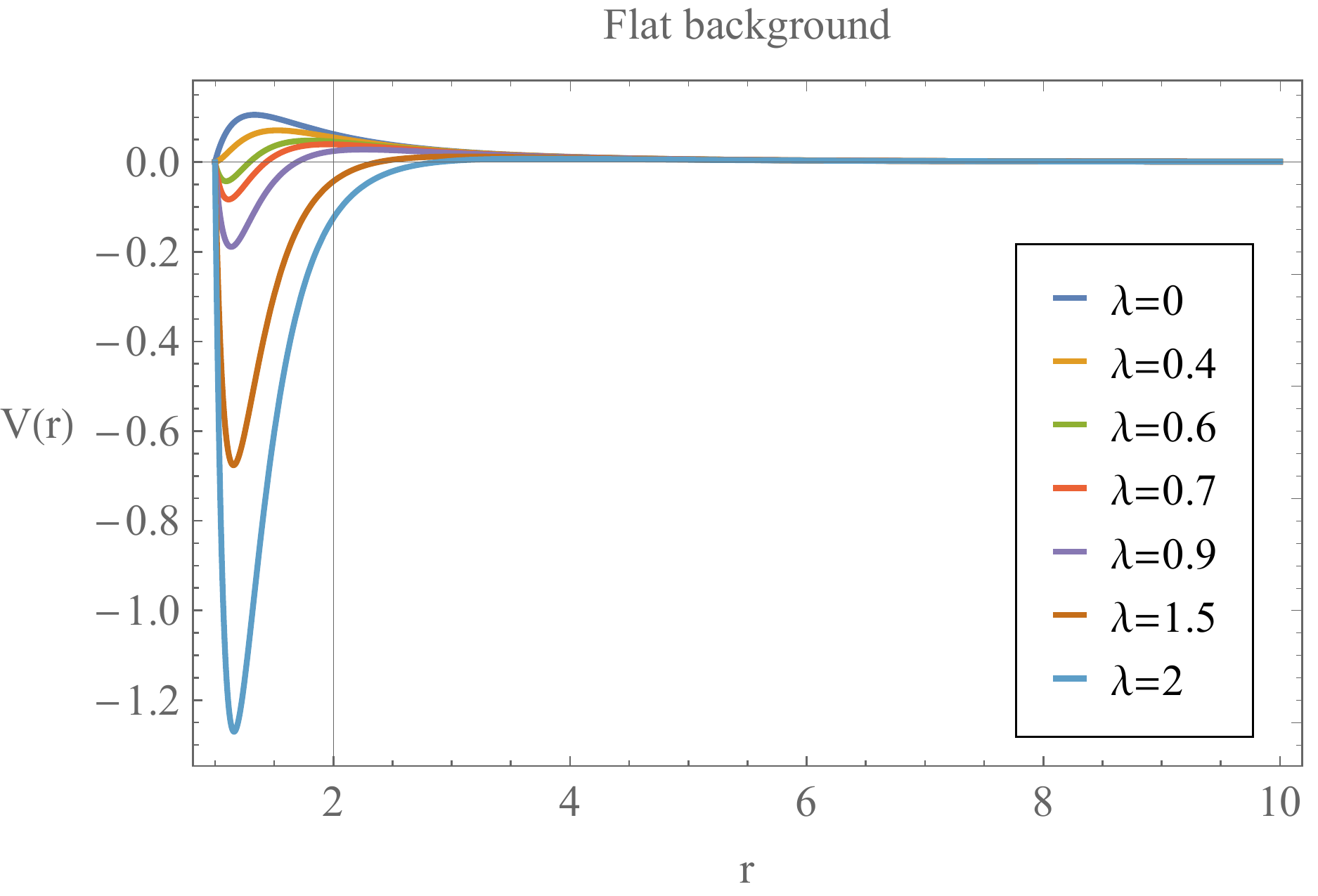}
  \caption{Left: The behaviour of the effective potential for the AdS black hole as the function of $r$ outside the horizon. Right: The effective potential for the flat case with $L\to\infty$ with the same value of parameters. Here we have set $r_h=1$ and $m_e^2=-2$ without loss of generalization.}\label{effpoten}}
\end{figure}
%%%%%%%
%%%%%
\begin{figure}[h]
\center{
  \includegraphics[scale=0.4]{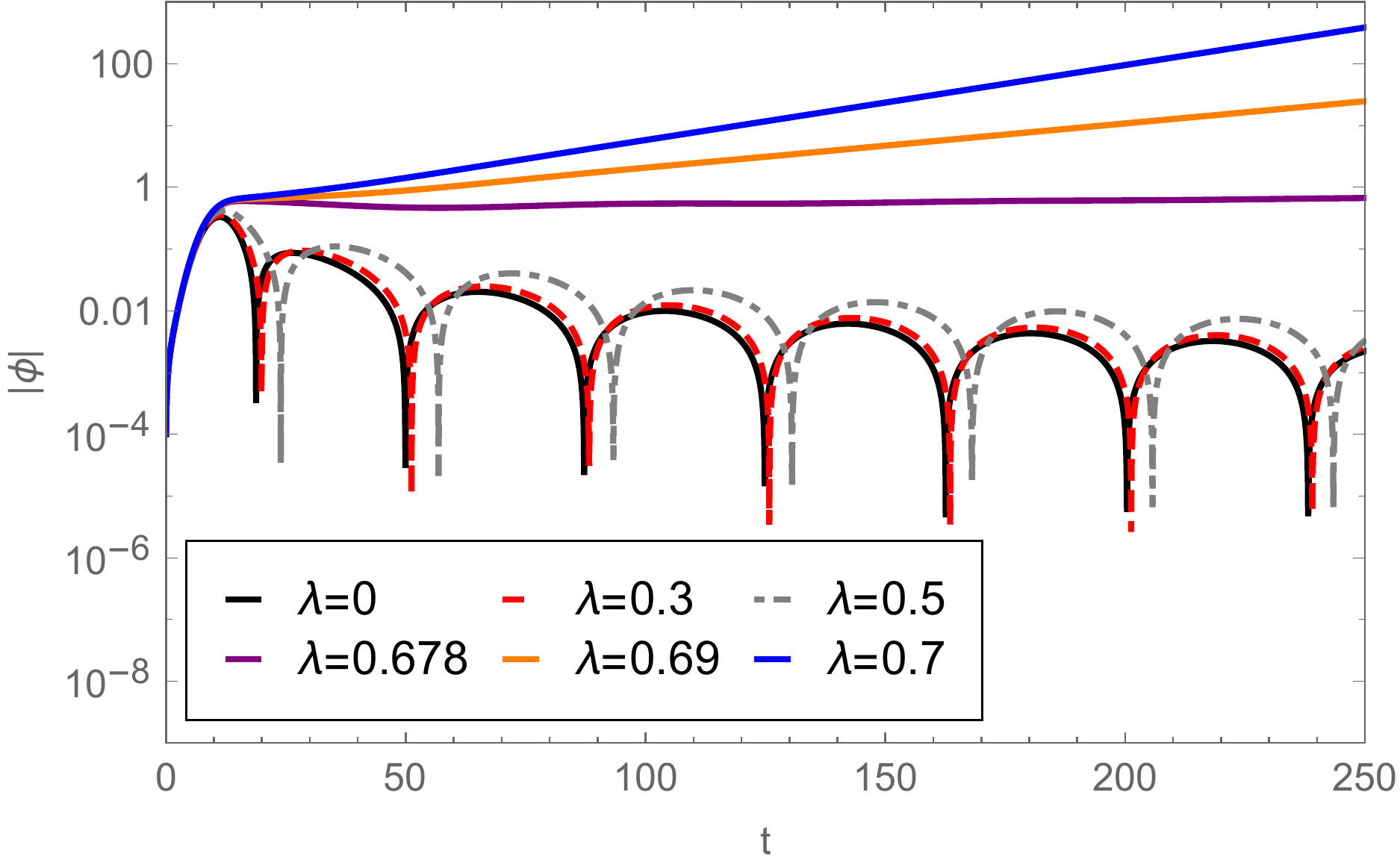}
  \caption{The time domain profile of the scalar field at fixed radial coordinate $r=1000$.}\label{tphi}}
\end{figure}

We exhibit the profile of the effective potential in the left panel of Fig. \ref{effpoten}. As is shown, when $\lambda$ increases to some critical value, there emerges the potential well. With the growing of the coupling, the potential well becomes deeper and larger. This behaviour indicates that there exists certain instability  of the scalar field. With more careful study, we find that the potential well emerges at the coupling around $\lambda \approx 0.6$. Meanwhile, in the right plot we show the effective potential for the flat Schwarzschild background, i.e. $L\to\infty$.
It is obvious that the potential well in AdS case is more larger and deeper. This implies that comparing to the flat case,  the instability brought by the GB coupling for the planar black hole in AdS case could occur easier.

Moreover, we obtain the time domain profile of the scalar field by solving the time dependent perturbation equation at the radial distance $r=1000$. The results are shown in the right panel of Fig. \ref{tphi}. Below a critical value of the coupling, the scalar field decays with time, which implies the stability of the scalar field. However, when the coupling parameter $\lambda$ increases over a critical value $\lambda\approx 0.678$, the scalar field grows up with time. This behaviour is consistent with the analysis of the effective potential.

\section{Scalarization analysis and holography in Einstein-scalar-Gauss-Bonnet Theory}\label{sectIII}

\subsection{Signal of scalarization in the probe limit}
From the analysis in above section, we know that the scalar field could be unstable from dynamical analysis. On the other hand, in asymptotical AdS space time, the massive, real scalar field possesses a classical instability when the effective mass is below the Breitenlohner-Freedam bound \cite{BFbound}, and in our study, it is $m_e^2<m_{BF}^2=-\frac{9}{4}$ where the effective mass $m_e$ will be defined later.

In order to peer the possible fate of the scalar field, we shall study the boundary behaviour of the scalar field induced by the GB coupling term in the fixed SAdS background geometry.
Then, the Klein-Gorgon equation for $\phi=\phi(r)$  under the background \eqref{background} is given by
\begin{eqnarray}
\phi''(r)+\left(\frac{2}{r}+\frac{g'(r)}{g(r)}\right)\phi'(r)- \frac{m^2}{g(r)}\phi(r)+\frac{\lambda^2}{2g(r)}{\cal R}^2_{GB}\frac{df(\phi)}{d\phi}= 0~, \label{scalareq}
\end{eqnarray}
where the Gauss-Bonnet term is evaluated in \eqref{geometries}.
%\begin{equation}
%{\cal R}^2_{GB}=\frac{4}{r^2}\left(g'(r)^2+g(r)g''(r)\right)~.
%\end{equation}
Near the horizon, the above equation implies
\begin{eqnarray}
\phi'-\dfrac{1}{3} \left(-3-m^2-18 \lambda^2 e^{-\beta \phi^2}\right)\phi=0~,\label{bdy-horizon}
\end{eqnarray}
while at the infinity their behaviour is
\begin{eqnarray}\label{boundary}
  \phi(r) = \frac{\phi_-}{r^{\Delta_-}}+\frac{\phi_+}{r^{\Delta_+}}~
\end{eqnarray}
where $\Delta_{\pm}=\frac{3\pm \sqrt{9+4m_e^2}}{2}$ and here we set $m_e^2=m^2L^2-12\frac{\lambda^2}{L^2}$.

\begin{figure}[h]
\center{
  \includegraphics[scale=0.4]{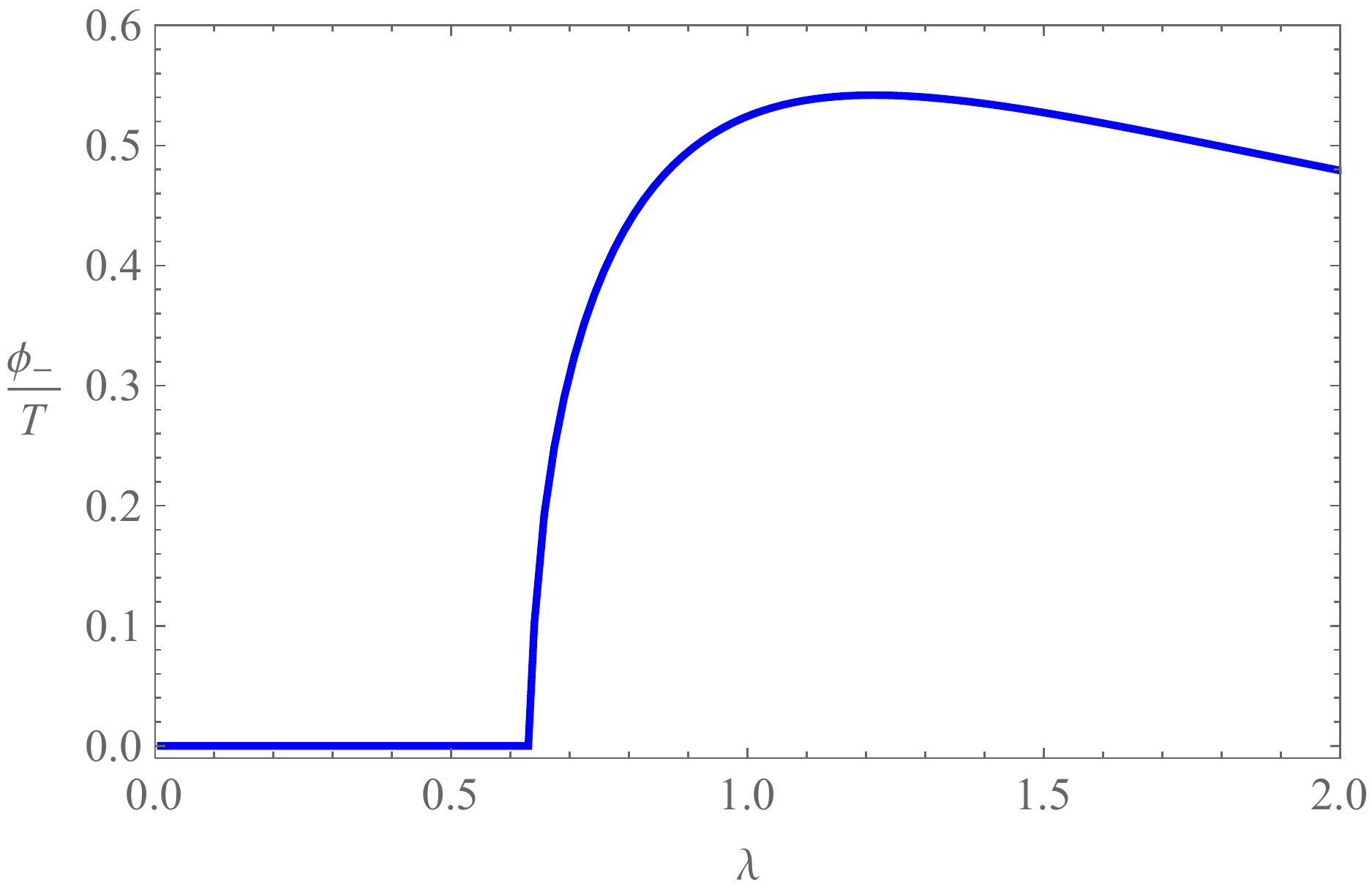}
  \caption{The boundary condensate parameter $\phi_-$ as a function of the Gauss-Bonnet coupling constant $\lambda$.\label{Dvslambda}}}
\end{figure}

As we mentioned previously, when $m_e^2>-9/4$, an instability would occur for the scalar field. We shall set $m_e^2=-2$  and numerically solve the equation \eqref{scalareq}. The behavior of $\phi_-$ v.s. the coupling strength is present in Fig. \ref{Dvslambda}. It is obvious that as the coupling parameter increases to a critical value, $\lambda_{c}\approx 0.64$, the $\phi_-$ emerges and enlarges rapidly.
%%%%%%%

\begin{figure}[h]
\center{
  \includegraphics[scale=0.4]{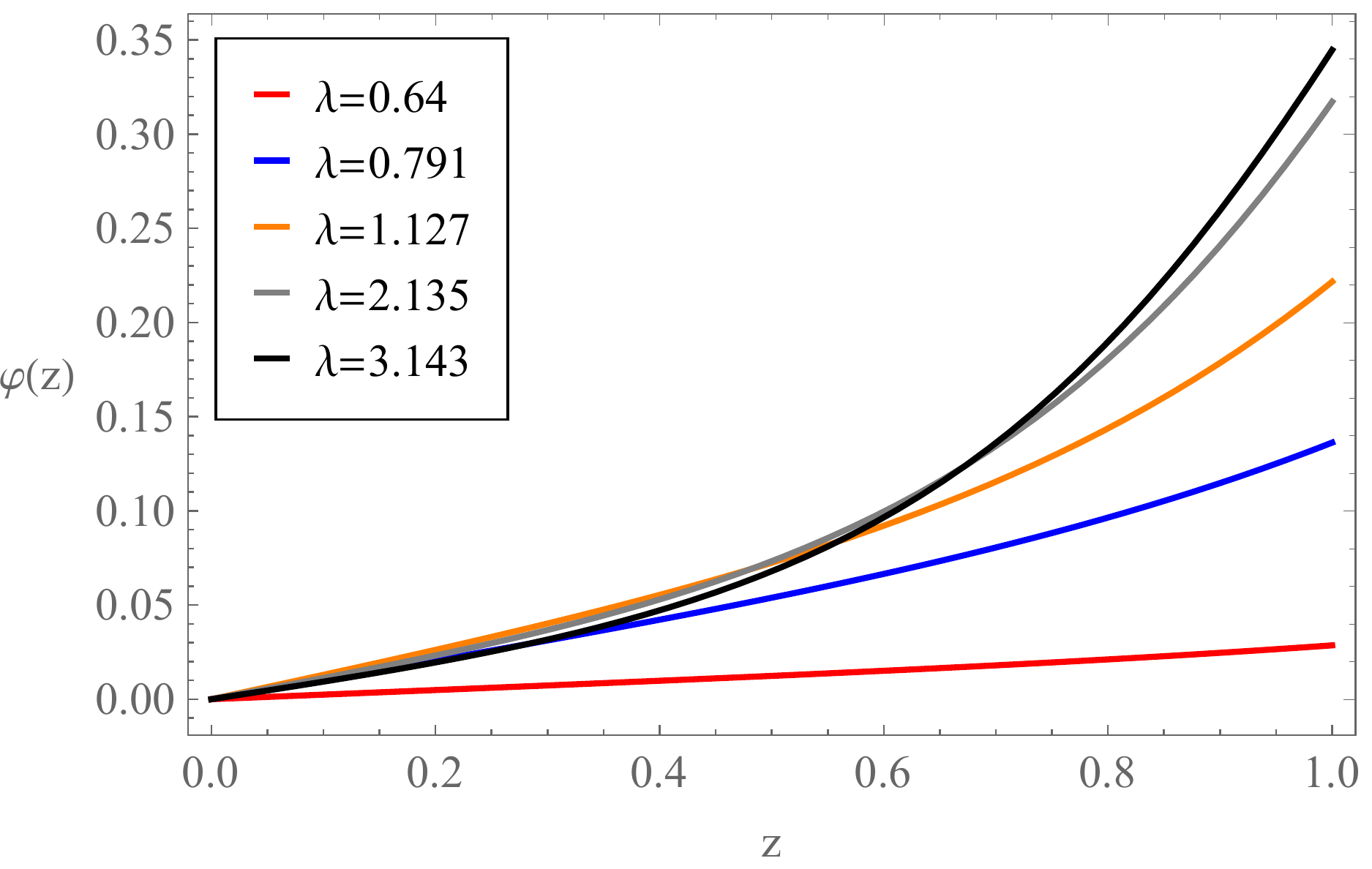}
  \caption{The profile of the scalar field as the function of the coordinate $z=\frac{r_h}{r}$ here we set the horizon $r_h=1$. We see that the scalar field condensates near the horizon contributing to the black hole hair for some intermediate $\lambda$. }\label{phivsz}}
\end{figure}

Moveover, we show the profile of the scalar field in Fig. \ref{phivsz}. It is obvious that the scalar field becomes nonzero when $\lambda$ is larger than the critical coupling. Different from the non-monotonic behavior of $\phi_-$, the scalar at the horizon monotonically increases as the coupling parameter becomes larger. Similar behaviour is also observed in \cite{Doneva_2018a}.
%%%%%%%%%%%%%%%%%%%%%%%%%%%%%%%%%%%%5

Since we are working in the probe limit, so the emergence of non-zero $\phi_-$ could be a strong signal of spontaneous scalarization of the theory. Next, we will explicitly construct the scalarized  hairy solutions by considering the backreaction of the scalar on the Schwarzschild AdS geometry.

\subsection{Scalarization with backreaction and the dual phase transition}
\subsubsection{Scalarized hairy black hole solution}

In this section, we will construct the hairy solution with the backreaction of the scalar field to the gravitational system. To this end, we consider the ansatz
\begin{equation}
	ds^2=\frac{1}{z^2}\left[-(1-z)p(z)U(z)dt^2+\frac{1}{(1-z)p(z)U(z)}dz^2+V(z)dx^2+V(z)dy^2\right]~,\label{eq-ansatz}
\end{equation}
where $z=r_h/r$, $p(z)=1+z+z^2$ and $U(z)$ and $V(z)$ are metric functions to be determined, thus the horizon is located at $z=1$ and the asymptotical boundary is at $z\to0$. Then the Klein-Gordon equation is
\begin{equation}\label{scalar}
    \phi''(z)+\left(\frac{p'(z)}{p(z)}+\frac{U'(z)}{U(z)}+\frac{V'(z)}{V(z)}-\frac{z-2}{z(z-1)}\right)\phi'(z)
    +\frac{1}{(z-1)z^2p(z)U(z)}\left(m^2\phi(z)+\frac{\lambda^2}{2}\frac{df(\phi)}{d\phi}{\cal R}^2_{GB}\right)=0~,
\end{equation}
and the non-vanishing components of Einstein equations are
\begin{eqnarray}\label{eintt}
    U'\left[2z(z-1)pV^2\left(zV'-2V\right)+\lambda^2\frac{df(\phi)}{d\phi} U M\right]+U[(z-1)z^2V\left(2pVV'-pV'^2\right)\nonumber \\
    -4z(z-1)V^3p'+2pV^3(-6+4z+(z-1)z^2\phi'^2)-2zpV^2((3z-4)V'-2z(z-1)V'')]\nonumber\\
    +\lambda^2\frac{df(\phi)}{d\phi}U^2[\frac{p'}{p}M+4(z-1)^2z^4p^2V'\phi'(2V''-V'^2)+8z(z-1)p^2V^3N
    \nonumber\\+2(z-1)z^3p^2VV'^2(N+4(z-1)\phi')-8(z-1)z^2p^2V^2(2z(z-1)\phi'V''+NV')]
    +2V^3(6-m^2\phi^2)~,
\end{eqnarray}
\begin{eqnarray}\label{leinzz}
    V'^2\left[(z-1)z^2pU+\frac{z^2}{4}X\right]+VV'[2(z-1)z^2(Up'+pU')-2z(3z-4)pU-zX]\nonumber\\ +V^2\big[12-2m^2\phi^2-4z(z-1)(Up'+pU')+X-2pU(6-4z+(z-1)z^2\phi'^2)\big]~,
\end{eqnarray}
\begin{eqnarray}\label{einxx}
	V''[-2(z-1)z^2pU+2\phi'\lambda^2\frac{df(\phi)}{d\phi}(2(z-1)(z-2)z^3p^2U^2-(z-1)^2z^4(p^2U^2)')]\nonumber\\
	+\frac{V'^2}{V}[(z-1)z^2pU+\frac{z^2}{12}X]+V'[2z(z-2)pU-2(z-1)z^2pU'-2Y]+V[-12+2m^2\phi^2\nonumber\\
	+2pU(2(3-z)-(z-1)z^2\phi'^2)+4z(z-2)Up'+2(z-1)z^2(2p'U'+Up'')\nonumber\\ +2zp(2(z-2)U'-z(z-1)U'')+\frac{4}{z}Y]~,
\end{eqnarray}
where
\begin{eqnarray}
	M&=&6(z-1)^2p^2V\phi'(4V^2-4zVV'+z^2V^2)~,\nonumber\\
	N&=&(z+2)\phi'+2(z-1)z\phi'',\nonumber\\
	X&=&12\phi'\lambda^2\frac{df(\phi)}{d\phi}[(z-1)^2z^2(p^2U^2)'-2z(z-1)(z-2)p^2U^2]~,\nonumber\\
	Y&=&\lambda^2\frac{df(\phi)}{d\phi}\{[(z-1)^2z^4\phi'(U^2p^2)']'-2(z-1)z^4\phi'(U^2p^2)'-Q\}~,\nonumber\\
	Q&=&2z^2p^2U^2[(z^2-2)\phi'+z(z^2-3z+2)\phi'']~.
\end{eqnarray}
It is easy to check that when $U(z)=V(z)=1$ and $\phi=0$, the metric \eqref{eq-ansatz} is a solution of the above system which is nothing but the SAdS background geometry.

The asymptotical $AdS_4$ of the metric requires the boundary condition $U(z=0)=U(z=0)=1$, and the scalar field behaves the same as \eqref{bdy-infity}, i.e., $\phi(z\to 0)=\phi_-z^{3- \sqrt{9+4m_e^2}/2}+\phi_+z^{3+ \sqrt{9+4m_e^2}/2}$. We shall set the model parameters the same as those in the probe limit and solve the above differential equations group numerically.

\begin{figure}[h]
\center{
  \includegraphics[scale=0.3]{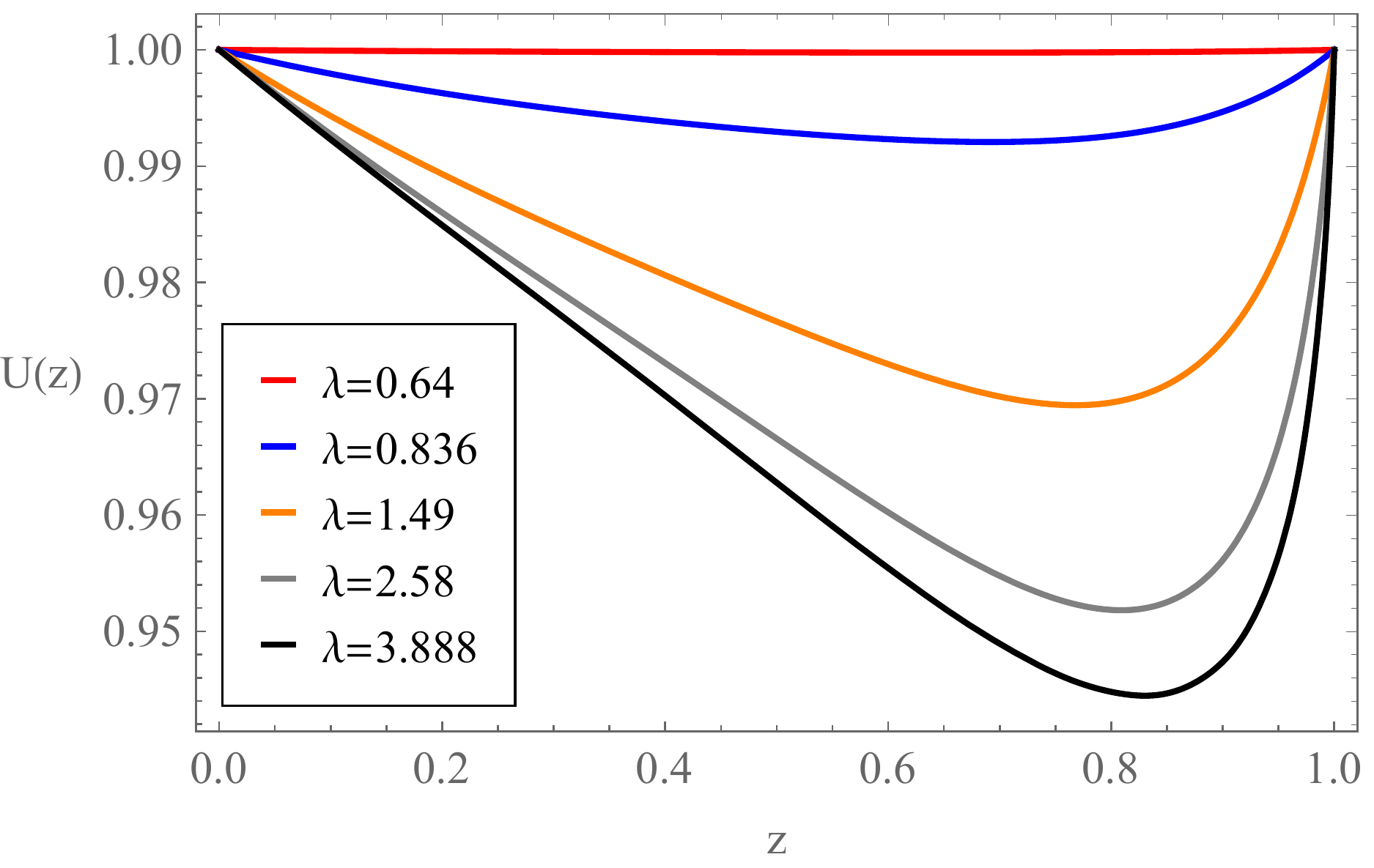}
  \includegraphics[scale=0.26]{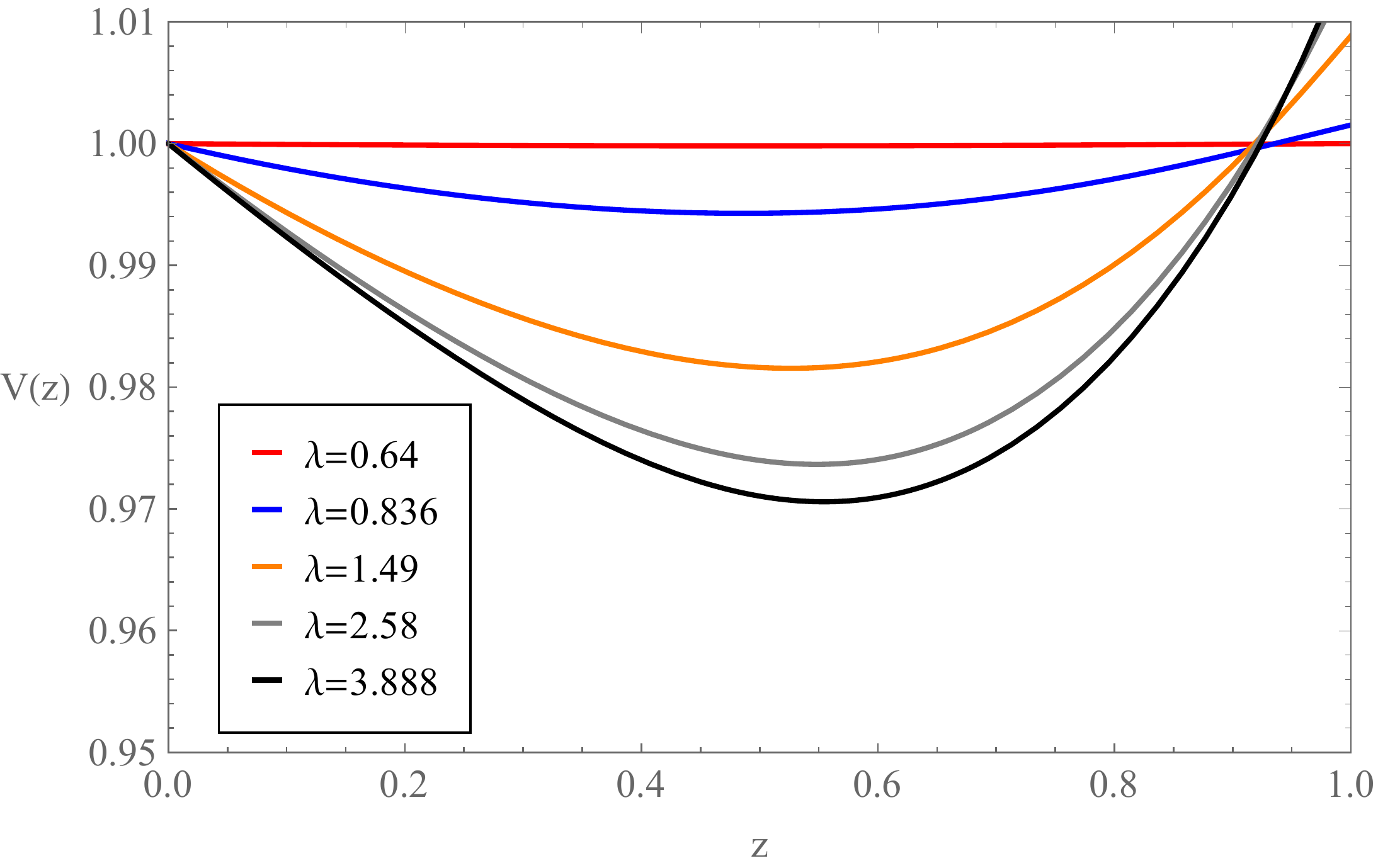}
  \includegraphics[scale=0.3]{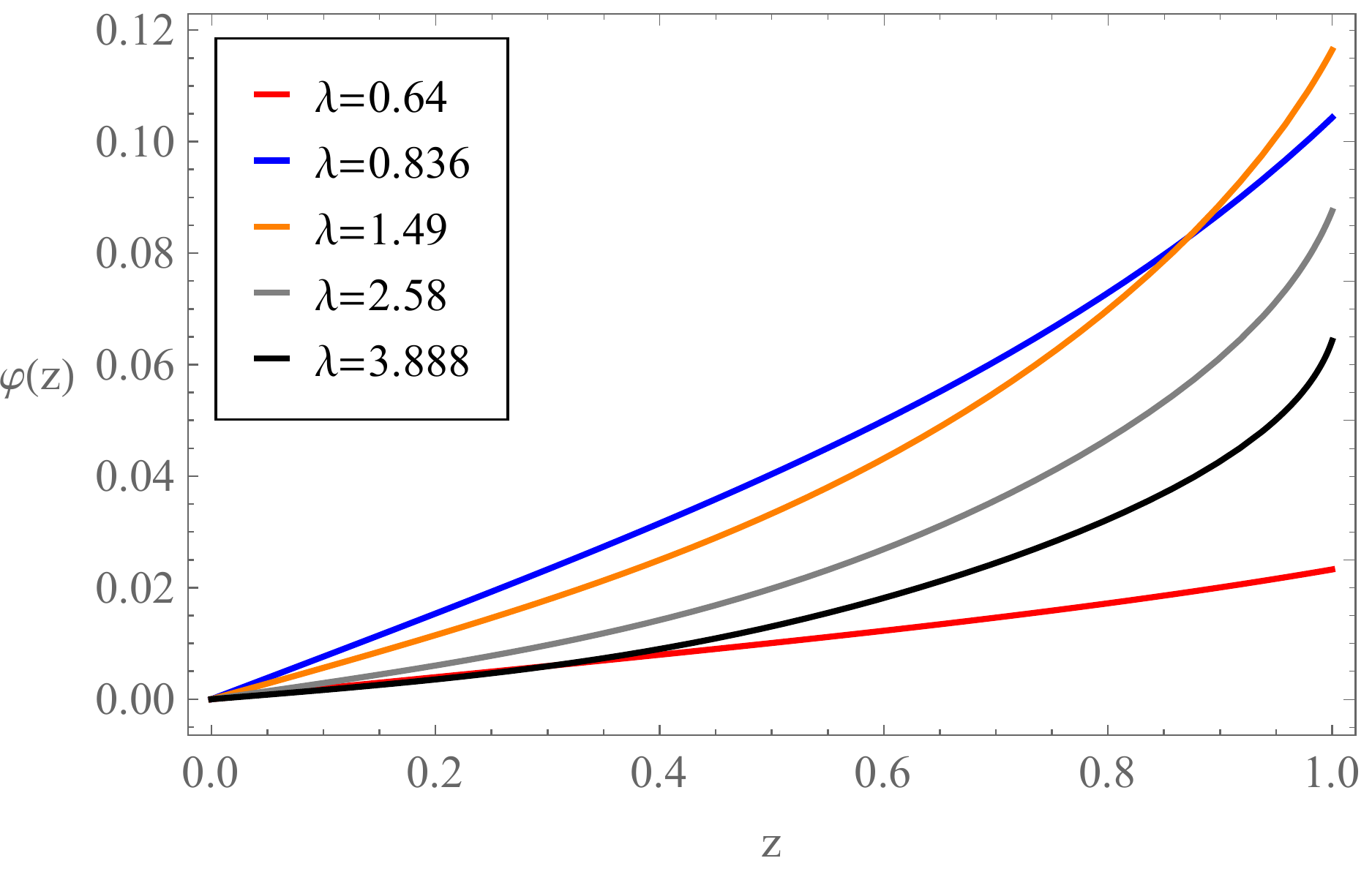}
  \caption{The profiles of the metric function $U,V$ and the scalar function $\phi$ of the hairy black hole as the function of the coordinate $z=\frac{r_h}{r}$.}\label{profile}}
\end{figure}

By tunneling the coupling parameter $\lambda$, we find the nonzero solution of the scalar hairy of black hole. The profiles of the metric functions $U(z)$, $V(z)$ and the non-vanishing scalar field $\phi(z)$ are shown in Fig. \ref{profile}. It is obvious that near the critical coupling, the $U(z)$ and $V(z)$ are close to unit because the scalar hair starts to form. As $\lambda$ increases, $U(z)$ and $V(z)$ shift from unit which means that a new hairy black hole solution forms. In the right plot for the scalar field, we see that as the coupling increases, the scalar field at the horizon increases first and then decreases.
%%%%%%%%%%%%%%%%%%%%%
%%%%%%%%%%%%%%%%%%
\begin{figure}[h]
\center{
  \includegraphics[scale=0.4]{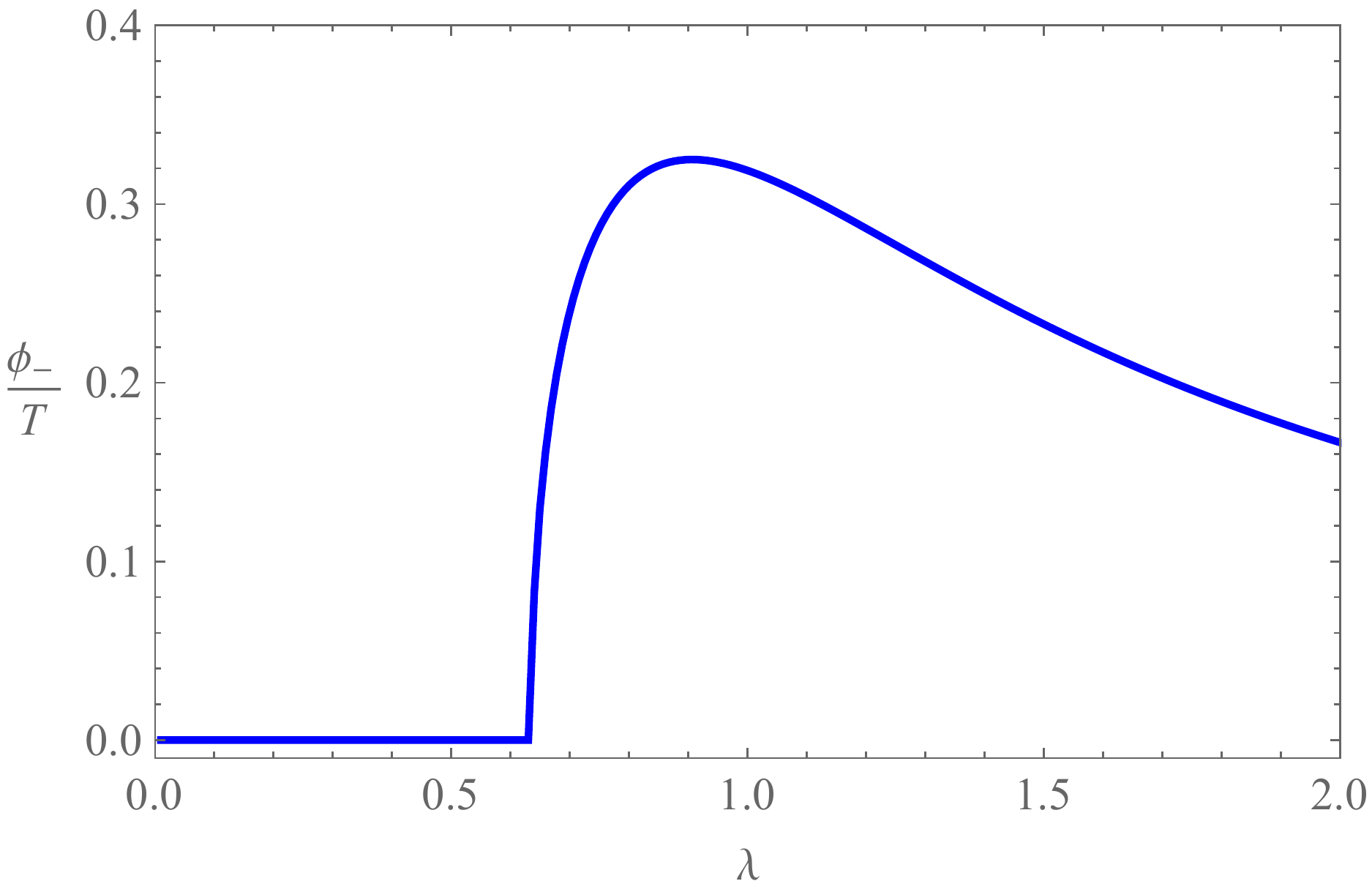}
  \caption{The boundary condensate parameter $\phi_-$ as a function of the coupling parameter $\lambda$. The condensate merges around $\lambda_{c}\approx 0.64$  which are the same as the results in the probe limit.}\label{BackDvslambda}}
\end{figure}

According to AdS/CFT duality, the scalar field is dual to a scalar operator in the boundary theory.  $\phi_-$ or $\phi_+$ in \eqref{bdy-infity} is dual to the source while the other is the vacuum expectation value (VEV) of the operator in term of the choice of the standard quantization or alternative one. Here we choose the coefficient of higher order term, $\phi_+$, as the source, i.e., $\phi_+=0$ and treat $\phi_-$ as vacuum expectation value in the
dual theory.
The VEV, $\phi_-$ as a function of the coupling parameter is plotted in Fig. \ref{BackDvslambda}. Similar to the case in the probe limit, as the coupling increases, $\phi_-$ becomes nonzero around  $0.64$, and it increases sharply to a maximum meaning a hairy black hole emerges.  $\phi_-$ would decrease as we further increase the coupling, this may be because the scalar field is too difficult to live around the black hole due to so strong gravity\footnote{Due to the nonlinearity, the numerics would break down for large coupling.}.

Our study shows that the GB coupling  destabilize the SAdS planar black hole and a hairy black hole could form.
As we already mentioned  the formation of the hairy black hole because of interaction between the scalar field and GB curvature correction is apparently different from that in the Abelian Higgs model \cite{Gubser:2005ih}. Then we shall discuss the differences both in the bulk gravity side and on the boundary CFT side.
On one hand, in the bulk gravity side, the scalarization occurs because the GB coupling term contributes to the effective mass of the scalar field, so that it is possible to be lower than the BF bound and cause instability even though only the gravitational force is involved. While in the Abelian Higgs model, the nonzero  $U(1)$ gauge field could suppress the effective mass.

The formation of a hairy black hole as a result of this instability was discussed in \cite{Gubser:2008px}, and the gravitational attraction and the electromagnetic repulsion are two competing forces in the system. On the other hand, from the viewpoint of the dual boundary theory, it is known that a black hole in the gravity side is holographically dual to a thermal state in the boundary field theory (CFT). Consequently, the formation of hairy black hole in the bulk can be interpreted as a certain phase transition characterizing  by non-vanishing VEV emerged in the boundary theory, and the GB coupling somehow mimics a  mechanism that leads to the phase transition. Since there is no symmetry in this setup, so the scalarization could be dual to certain quantum critical phase transition  which usually does not accompany with breaking of a symmetry.  Similar phase transition at finite temperature was holographically studied by adding the new coupling between the scalar field and the Weyl curvature \cite{Katz:2014rla,Myers:2016wsu}.
While, the formation of a hairy black hole of Abelian Higgs model is dual to holographic superconductor phase transition\cite{Hartnoll:2008kx,Hartnoll:2008vx}. Above a critical temperature the VEV is zero and the system is  dual to a normal state, while below the critical temperature, the VEV becomes nonzero and the system is dual to a holographic superconducting state.

\subsubsection{Holographic entanglement entropy as a probe}

As is known that entanglement entropy(EE)  is an important physical quantum in quantum field theory. It plays a crucial role in holographic framework, especially in the further understanding of quantum gravity and phase transition  physics. It has been proposed that in holographic framework, the EE for a subregion
on the dual boundary is proportional to the minimal surface in the bulk geometry, for which
is being called the Hubeny Rangamani-Takayanagi (HRT) surface\cite{Ryu:2006bv}, i.e., holographic entanglement entropy (HEE) is a well-known geometrical description of EE in the dual boundary theory.  One of the most important applications of HEE and the studies of HRT surfaces is to diagnose and study various phase transitions, for instance holographic  superconducting phase transition \cite{Albash:2012pd,Cai:2012nm,Kuang:2014kha,Guo:2019vni}, quantum phase transition\cite{Ling:2015dma,Pakman:2008ui}, confinement/disconfinement phase transition in QCD\cite{Klebanov:2007ws,Zhang:2016rcm} and thermodynamic phase transition\cite{Zeng:2016fsb}  and so on.

 In this subsection, we will probe the scalarized process brought by the GB coupling by calculating the HEE of the dual theory,  which is one of the most important characteristic scales of the boundary theory. Especially, we expect that the HEE would be a good probe of scalarization, in the sense  it would characterize the phase transition in the dual boundary theory.

We shall apply Ryu-Takayanagi proposal \cite{Ryu:2006bv} to calculate HEE of the sector. To this end, we consider the subsystem $A$ with a straight strip geometry described by $-\frac{l}{2}\leq x \leq\frac{l}{2}\,, 0\leq y \leq L$, where $l$ is  the size of $A$ and $L$ is a regulator which can be set to be infinity.  Ryu and Takayanagi proposed in \cite{Ryu:2006bv} that the HEE $S_A$ is determined by the radial minimal extended surface $\gamma_A$ bounded by $A$ via
\begin{equation}
	S_A=\frac{Area(\gamma_A)}{4G_N}~,
\end{equation}
in Einstein gravity. Then in \cite{Dong:2013qoa}, it was proposed a general formula for HEE in higher derivative gravity
\begin{eqnarray}\label{HEE0}
S_A&&=-2\pi \int_{\Xi}d^2x\sqrt{h}\frac{\partial L}{\partial R_{\mu\nu\rho\sigma}}\varepsilon_{\mu\nu}\varepsilon_{\rho\sigma}+\mathrm{Anomaly~~term}\nonumber\\
&&=-2\pi \int_{\Xi} d^2x\sqrt{h}\left(\frac{\partial L}{\partial R_{\mu\rho\nu\sigma}}\varepsilon_{\mu\rho}\varepsilon_{\nu\sigma}-\sum_\alpha\left(\frac{\partial^2 L}
{\partial R_{\mu_{1}\rho_{1}\nu_{1}\sigma_{1}}\partial R_{\mu_{2}\rho_{2}\nu_{2}\sigma_{2}}}\right)_{\alpha}\frac{2K_{\lambda_1\rho_1\sigma_1}K_{\lambda_2\rho_2\sigma_2}}{q_{\alpha+1}}\right.\nonumber\\
&&\left.\times[(n_{\mu_1\mu_2}n_{\nu_1\nu_2}-\varepsilon_{\mu_1\mu_2}\varepsilon_{\nu_1\nu_2})n^{\lambda_1\lambda_2}
+(n_{\mu_1\mu_2}\varepsilon_{\nu_1\nu_2}+\varepsilon_{\mu_1\mu_2}n_{\nu_1\nu_2})\varepsilon^{\lambda_1\lambda_2}]\right),
\end{eqnarray}
which  includes the Wald entropy as the leading term and the anomaly term of HEE corrected from extrinsic
curvature. Here $q_{\alpha}$ is treated as `anomaly coefficients ', $L$ is the Lagrangian density of action shown in \eqref{action} for neutral case, $h$ is
determinant of the induced metric on the extended surface $\Xi$ which minimizes the functional $S_A$.
%%%%%%%%%%
\begin{figure}[h]
\center{
  \includegraphics[scale=0.4]{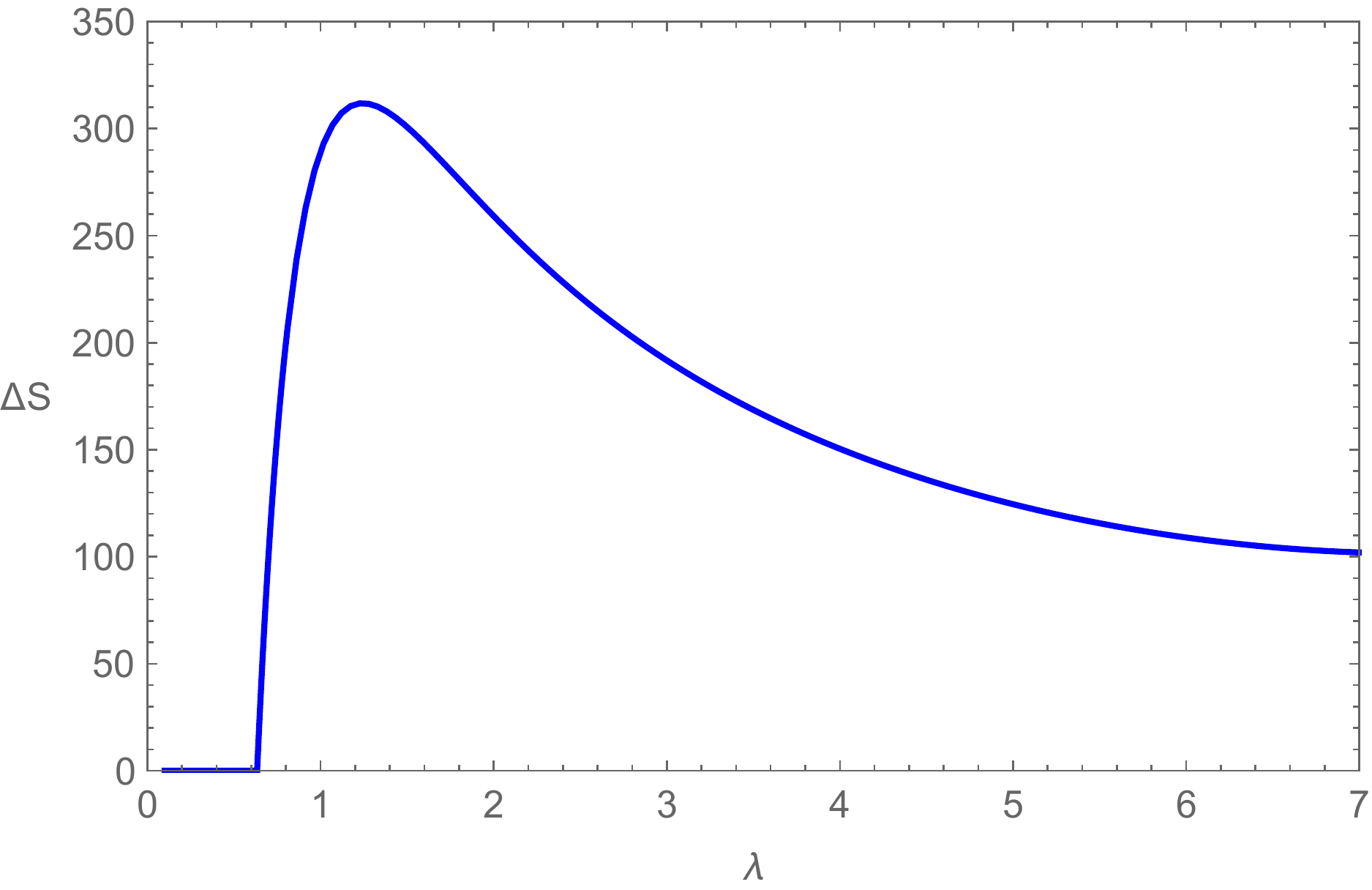}
  \caption{The difference of holographic entanglement entropy $\Delta S=S_A-S_{A0}$ as a function of the GB coupling.}\label{entan}}
\end{figure}
%%%%%%%%%%
In our model, due to the existence of the Gauss-Bonnet coupling term, we shall employ the above proposal to compute HEE for the dual theory, which is evaluated as
\begin{equation}
S_A=\frac{1}{4}\int_\Xi dx^2\sqrt{h}\left[1-f(\phi)\left(2R-4(R_a^a-\frac{1}{2}K_aK^a)
+2(R^{ab}R_{ab}-K_{aij}K^{aij})\right)\right].
\end{equation}
where $K_{aij}$ is the extrinsic curvature tensor and $K_a$ is defined as $K_a\equiv K_{aij}h^{ij}$.
It is noticed that in the above expression, the terms explicitly relative with the extrinsic curvature  are the contribution from anomaly brought by  the GB coupling,
while the remaining terms stem from the Wald entropy in the proposal \eqref{HEE0}.
Before we exhibit the numerical result of $S_A$, we shall refer to \cite{Dong:2013qoa} and briefly explain the notations in the above two formulas.
The Greek letters $\mu,\nu,\cdot\cdot\cdot$ are indices of 4- dimensional bulk geometry, and $i,j,\cdot\cdot\cdot$ are indices of 2-dimensional  extended surface $\Xi$ while   the Latin letters $a,b,\cdot\cdot\cdot$ are as  indices of 2-dimensional space orthogonal to $\Xi$. In terms of two orthogonal unit vectors $n_{\mu}^{(a)}$, we define $n_{\mu\nu}=n_{\mu}^{(a)}n_{\nu}^{(b)}G_{ab} $ which project to the induced 2-dimensional metric $G_{ab}$ in the $x^a$ directions.
Then the tensor $\varepsilon_{\mu\nu}$ could be constructed as $\varepsilon_{\mu\nu}=n^{(a)}_\mu n^{(b)}_\nu \varepsilon_{ab}$,
where $\varepsilon_{ab}$ is the usual Levi-Civita tensor and $\varepsilon_{\mu\nu}$ is nothing but the usual Levi-Civita tensor  in the two orthogonal directions with all other components vanishing.

When the coupling parameter $\lambda$ is smaller than the critical value, the SAdS black hole is the physically favorable solution and the HEE $S_{A0}$ is a constant which is independent of the coupling. When the $\lambda$ increases to be larger than the critical value, the scalarized hairy solution emerges and the HEE starts to grow.
This behaviour is present in Fig. \ref{entan}, where we show the relation between the difference $\Delta S=S_A-S_{A0}$ and the coupling parameter $\lambda$. Firstly, the critical coupling $\lambda_{c}\approx 0.64$, which agrees with that we obtained in previous study, can also be read from the jumping of HEE. Moreover, it is obvious that the HEE of the hairy solution is larger than the SAdS solution.   Since HEE is a measure of degree of freedom in a system, the scalar field should introduce new degree of freedom into the boundary theory dual to hairy black hole, causing the HEE increase after scalarization.
It is worthwhile to point out that this behavior of the HEE is quite different with that in holographic superconductors, where the hairy superconducting state usually has less HEE than the normal state because the emergence of cooper pairs in superconduting state reduces the degree of freedom of the system, see for example
\cite{Albash:2012pd,Cai:2012nm,Kuang:2014kha,Das:2017gjy,Peng:2014ira,Yao:2014fwa,Dey:2014voa,Romero-Bermudez:2015bma,
Peng:2015yaa,Momeni:2015iea,Guo:2019vni} and therein. Along with the different mechanisms as we studied in last subsection, this is another different features  between the phase transitions of scalarization and holographic superconductor. Deep physical essence of those and more differences between them deserve further efforts.

Moreover, scalarization means that at small distances $r>r_h$ there is a formation of halo of matter. This is a dynamical process and the only force is the gravitational force. Fig. \ref{entan} shows that the HEE first grows from $\lambda_c$ and then decreases until it is stabilized. This indicates that the black hole acquired hair even though scalar field goes inside the background black hole horizon  until the black hole is stabilized.  Since this is a small scale process, on the dual boundary can only be described as a quantum physics effect. From this perspective, we can again argue that the sclarization we discuss could correspond to a certain quantum phase transition, however, more effort should be made for deep physics in this direction.

\section{Holographic phase transition in Einstein-scalar-Gauss-Bonnet Theory in the presence of an electromagnetic field}\label{sectIV}
In this section, we will add the electromagnetic field into the Einstein-Scalar-Gauss-Bonnet theory and then the scalar field becomes charged.  According to the instability analysis in section \ref{secII}, the two different scalarization mechanisms accommodate a wider and deeper effective mass,  so this part shall show us a picture on their combined effect on the formation of hairy black holes. We shall investigate the  charged scalar field condensation in the probe limit in which the matter fields will not backreact into gravity. So the background metric is the  Schwarzschild-AdS planar black hole shown in \eqref{background} and \eqref{backgroundA},
%\begin{equation}
%ds^2=-g(r)dt^2+\frac{dr^2}{g(r)}+r^2(dx^2+dy^2)~,
%\end{equation}
%with $g(r)=\frac{r^2}{L^2}-\frac{M}{r}$
and $L$ and $M$ determines the Hawking temperature of the black hole
\begin{equation}
T=\frac{3M^{1/3}}{4\pi L^{4/3}}~.
\end{equation}

We expect a phase transition to occur  at certain critical temperature of  black hole and this  process according to gauge/gravity duality corresponds to  a holographic superconducting phase transition \cite{Hartnoll:2008vx,Hartnoll:2008kx}.

\subsection{Holographic superconducting condensation}
We then start from the theory with the action\eqref{action}.  We set $\phi=\phi(r)$ and $A=A_t(r)dt$, so the equations of scalar and electromagnetic field are shown as
\begin{eqnarray}
  A_t''(r)+\frac{2}{r}A_t'(r)-\frac{2q^2\phi(r)^2}{g(r)}A_t(r) = 0 \label{eq-at}~,\\
  \phi''(r)+\left(\frac{2}{r}+\frac{g'(r)}{g(r)}\right)\phi'(r)+ \frac{q^2 A_t(r)^2-m^2g(r)}{g(r)^2}\phi(r)+\frac{{\cal R}^2_{GB}}{2g(r)}\frac{df(\phi)}{d\phi}= 0~,\label{eq-phi}
\end{eqnarray}
where the Gauss-Bonnet term is evaluated in \eqref{geometries}.
%\begin{equation}
%{\cal R}^2_{GB}=\frac{4}{r^2}\left(g'(r)^2+g(r)g''(r)\right)~.
%\end{equation}
Near the horizon, the scalar and electromagnetic field are regular so at the horizon $r=r_h$, for $A_t(r)$ to have finite norm, $A_t(r_h)=0$, and Eq.~(\ref{eq-phi}) then implies
\begin{eqnarray}
\phi'(r_h)=\dfrac{L^2}{3r_h} \left(m^2-\dfrac{18 \lambda^2 e^{-6 \phi(r_h)^2}}{L^4}\right)\phi(r_h)~,\label{bdy-horizon}
\end{eqnarray}
while at the infinity their behaviour are
\begin{eqnarray}
  \phi(r) = \frac{\phi_-}{r^{\Delta_-}}+\frac{\phi_+}{r^{\Delta_+}}~, ~~~ A_t(r) =\mu-\frac{\rho}{r}~,\label{bdy-infity}
\end{eqnarray}
where $\Delta_{\pm}=\frac{3\pm \sqrt{9+4m_e^2}}{2}$ with $m_e^2=m^2-12\lambda^2$.
%%%%%%%%%%%
\begin{figure}[h]
    \centering
    \includegraphics[scale=0.5]{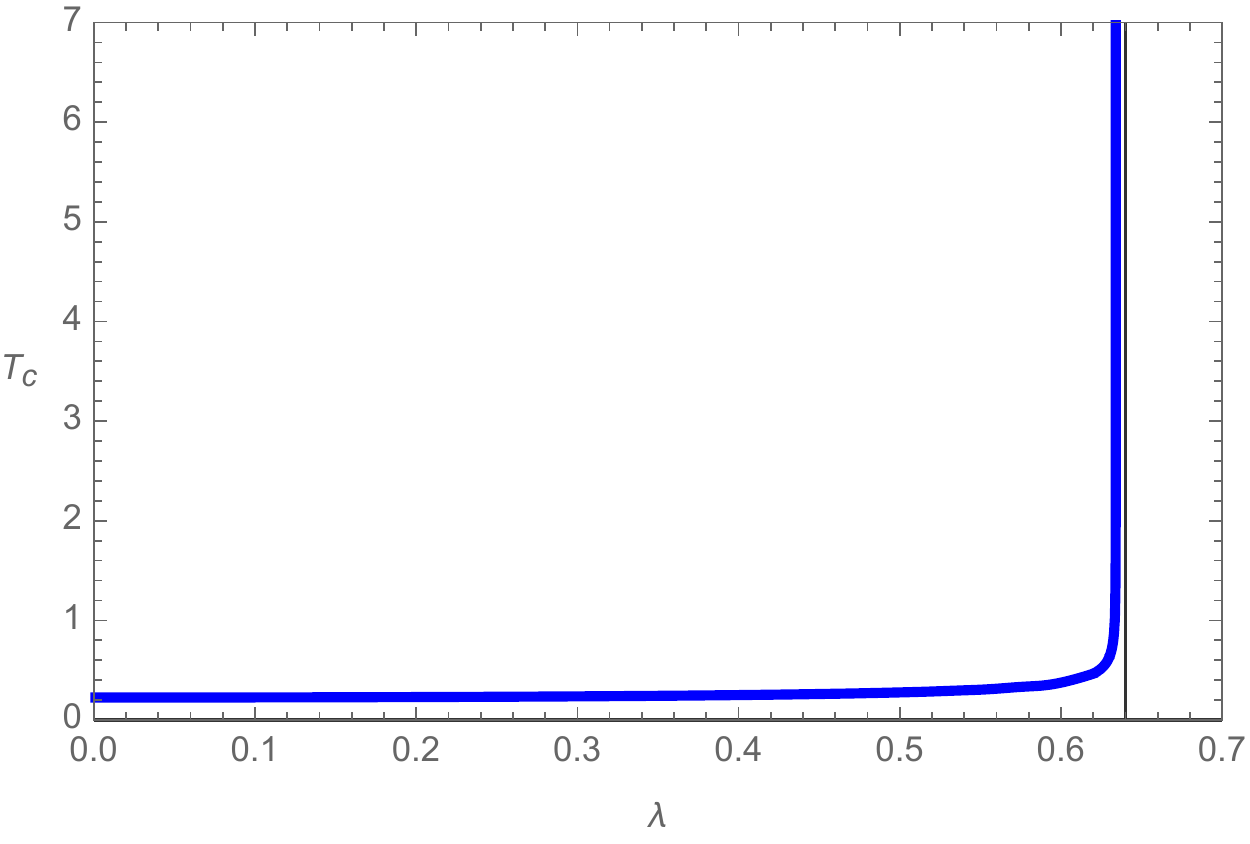}\hspace{0.5cm}
    \includegraphics[scale=0.53]{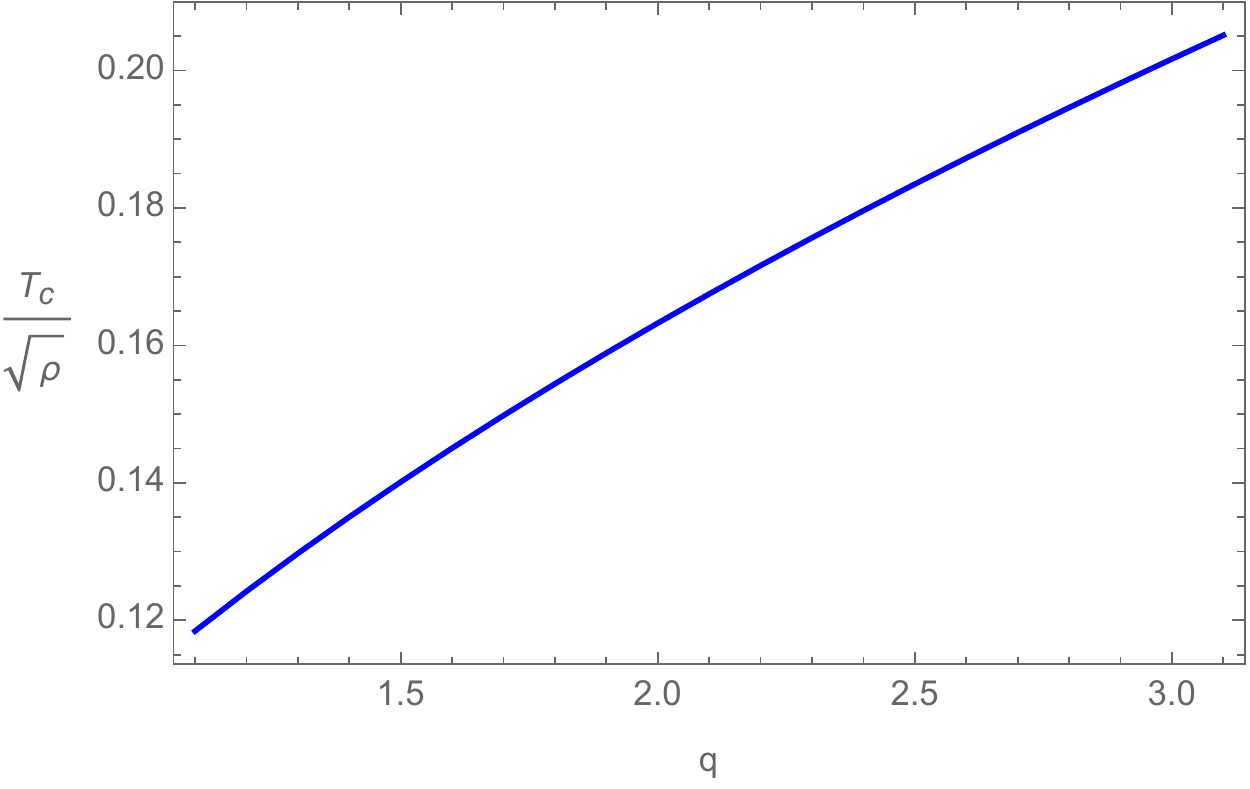}
    \caption{Left: the phase diagram  $\lambda-T_c$. Right: The phase diagram  $q-T_c$.}
    \label{criticalT}
\end{figure}
%%%%%%%%%%%%%%%%%%%%%%%%%%%%%%%%%%%%%
\begin{figure}[h]
    \centering
    \includegraphics[scale=0.4]{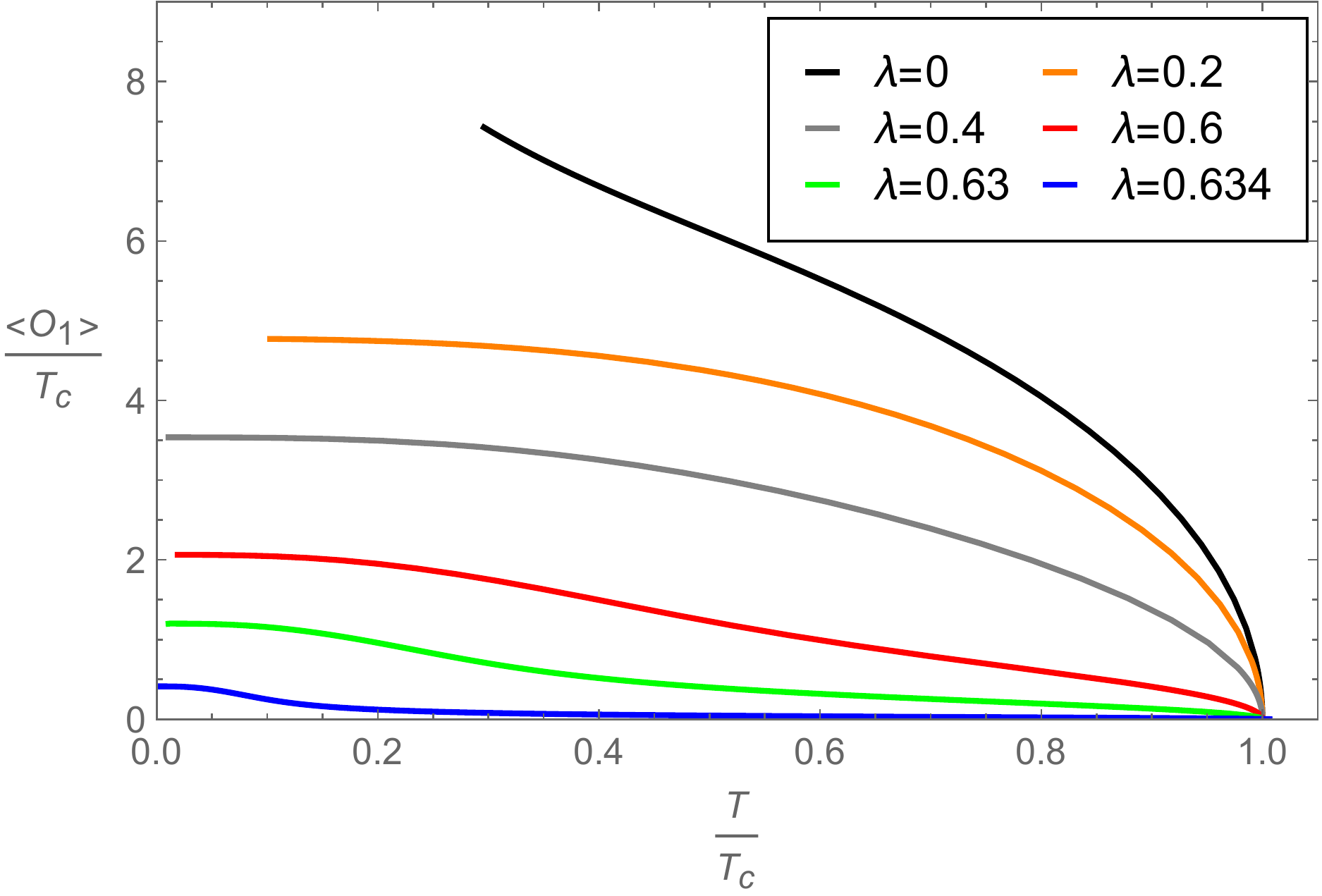}
    \caption{The condensation of scalar field as the function of the temperature $T/T_c$ with different couplings.}
    \label{condensate}
\end{figure}

By setting $L=M=q=1,  m_e^2L=-2$, we solve the above equation via the boundary conditions.
We obtain a phase diagram $\lambda-T_c$ in the left plot of Fig. \ref{criticalT}.  $T_c$ is the critical temperature of the holographic superconducting phase transition, above which the normal black hole is physically stable while below it the black hole is stabilized in a superconductiong state  with non-vanishing $\phi_-$. In this figure we observe that the critical temperature $T_c$  first slightly increases as the GB coupling $\lambda$ is increasing. When the coupling goes to a critical value $\lambda_{cc}\approx 0.6339$, $T_c$  increases dramatically and then becomes divergent. It implies that the holographic superconducting phase transition only can occur when $\lambda<\lambda_{cc}$ and when $\lambda\geq\lambda_{cc}$, a hairy black hole does not form in the gravity sector while on the boundary there is no any  holographic superconducting phase.

This result is very interesting. It indicates that as the GB coupling $\lambda$ becomes larger than a critical value, the gravitational attraction
from the GB high curvature term becomes stronger and the formation of the scalarized black hole is not possible. A similar effect was observed in \cite{Pan:2009xa} as well as \cite{Gregory:2009fj,Li:2011xja,Kuang:2010jc} and therein. It was found that  the strong curvature effects outside the horizon of a five-dimensional Gauss-Bonnet-AdS black hole, the holographic superconducting  mechanism is less effective as the GB coupling is increased.
%In the next section we will study if this behaviour is still valid in the case of the scalar field backreacting with the background black hole.

It is noticed that in the limit of $\lambda=0$, from the equations \eqref{eq-at}-\eqref{eq-phi} we recover the $s$-wave superconductor model \cite{Hartnoll:2008vx}. In the right plot of Fig. \ref{criticalT} we depict the dependence of the critical temperature on the charge of the scalar field. It can be seen that as the charge is increased the critical temperature also increases. With different couplings, we also study the condensation of the scalar field below the corresponding critical temperature, and the results are shown in Fig. \ref{condensate} remaining however smaller than its critical value $\lambda_{cc}$. As the GB coupling increases, the condensation gap is suppressed, meaning the copper pairs are less in the dual boundary theory. We shall verify this phenomena by studying the conductivity in the next subsection.

Combining our observers in neutral and charged cases, let us figure out the possible scalarized picture in Einstein-Scalar-Gauss-Bonnet gravity with a negative cosmological constant. Above certain temperature, only the GB coupling could play the role in the formation of scalar hair which is dual to a certain quantum phase transition in the boundary theory. While the temperature becomes lower than a critical value, the holographic superconducting condensation participates and we have the combined stronger effects on the formation of hairy holes.

\subsection{Optical conductivity}

To compute the conductivity in the dual CFT as a function of frequency we need to solve the Maxwell equation for fluctuations of the vector
potential $A_x$. We will calculate the conductivity in the probe limit with a charged scalar field in the background of Schwarzschild-AdS black hole.  The Maxwell equation at zero spatial momentum and with a time dependence of the form $e^{-i \omega t}$ gives
\begin{eqnarray}
	 A_x''(r)+\frac{g'(r)}{g(r)}A_x'(r)+\left(\frac{\omega^2}{g(r)^2}-\frac{2q^2\phi(r)^2}{g(r)} \right)A_x(r)=0~.
\end{eqnarray}

We will solve the perturbed Maxwell equation with ingoing wave boundary conditions at the horizon, i.e.,
 $A_x  \propto g(r)^{-i\omega/3r_h}\mid_{r\to r_h}$. The asymptotic behaviour of the Maxwell field at large radius  is  $A_x = A_x^{(0)} + \frac{A_x^{(1)}}{r} + \cdots$. Then, according to AdS/CFT dictionary,  the dual source and expectation value for the current are given by $ A_x = A_x^{(0)} $ and $ \langle J_x \rangle = A_x^{(1)}$, respectively. Then the conductivity is given by the Ohms law
\begin{equation}
\sigma(\omega)= - \frac{i A_x^{(1)}}{\omega A_x^{(0)}}~.
\end{equation}

\begin{figure}[h]
\center{
\includegraphics[scale=0.4]{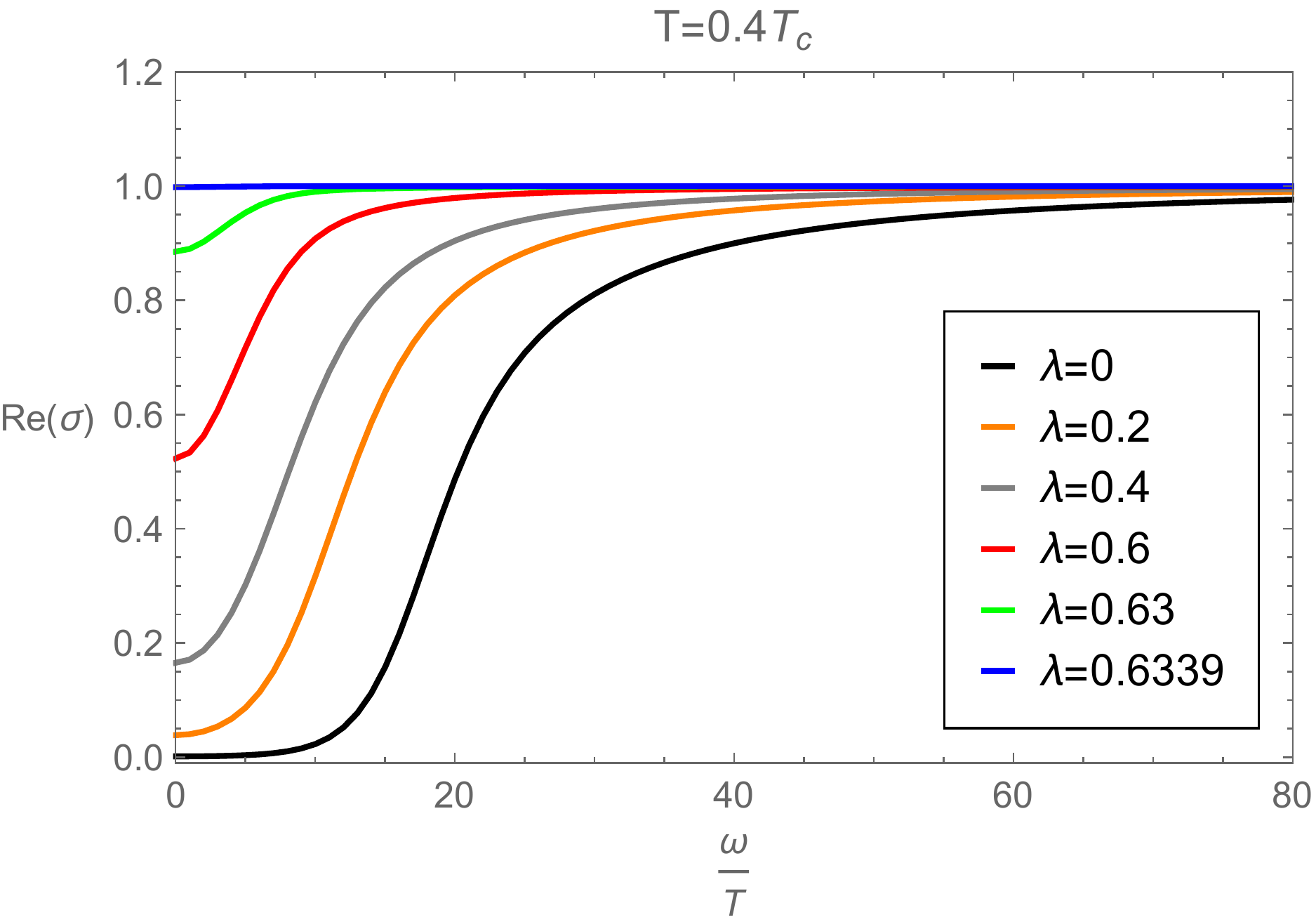}\hspace{0.5cm}
\includegraphics[scale=0.4]{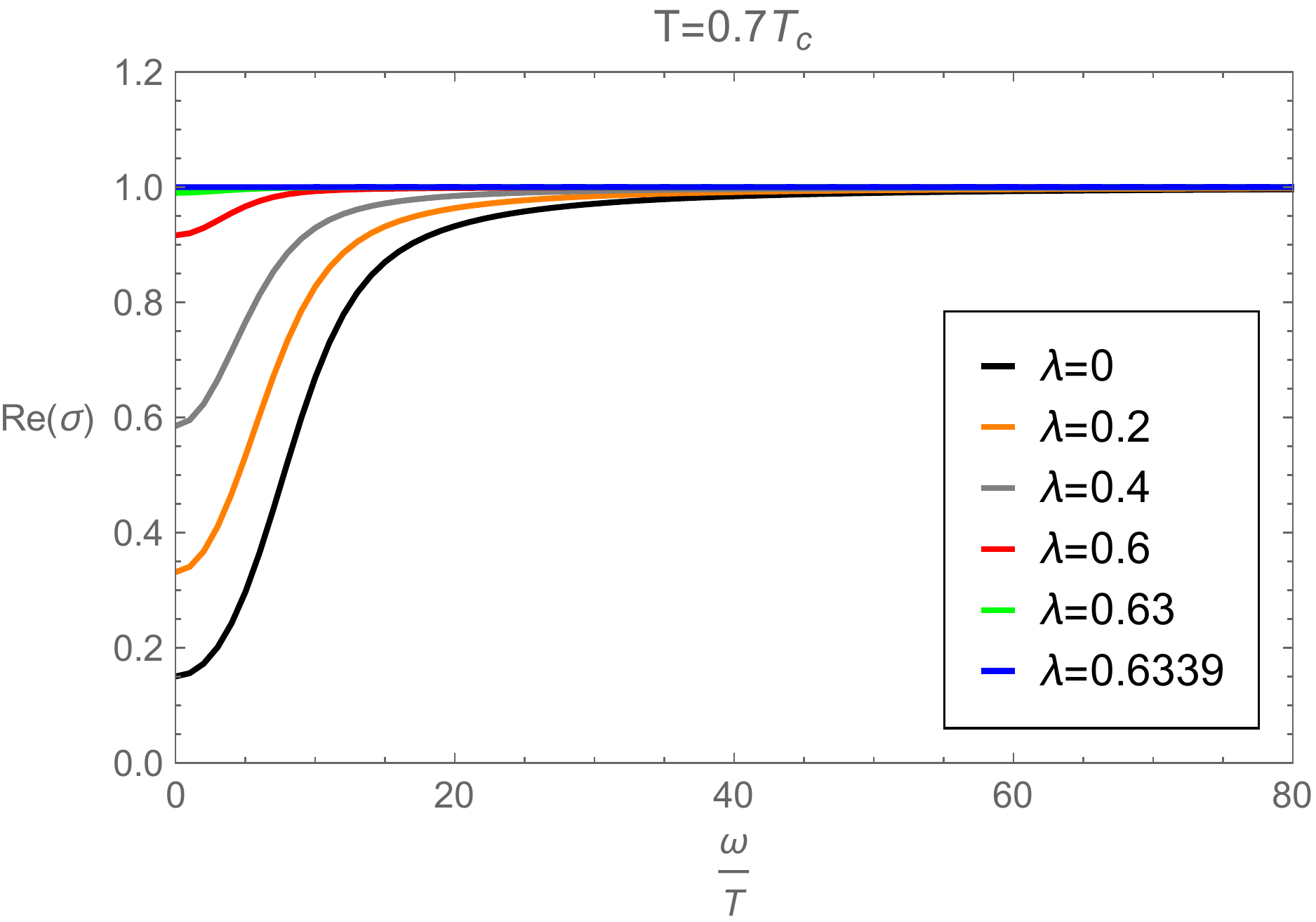}
\caption{The real part of optical conductivity, $Re(\sigma)$, as the function of $\frac{\omega}{T}$ at the temperature $0.4T_c$ and $0.7T_c$, respectively. } \label{fig-conduct}}
\end{figure}

The results of conductivity are shown in Fig. \ref{fig-conduct}. We see that as the coupling becomes stronger, the real part of $\sigma$ increases to unit for all frequency, which means that the conductivity is weaker and finally the system is dual to a metal at fixed temperature.

We can also  explore the behaviour of conductivity at very low frequency. When $T<T_c$, the real part of the conductivity present a delta function at zero frequency and the imaginary part has a pole, which is attributed to the following Kramers-Kronig (KK) relations
%%%%%%%
\begin{equation}\label{eq:KK}
\mathrm{Im}[\sigma(\omega)]=-\frac{1}{\pi}\mathcal{P}\int_{-\infty}^{\infty}\frac{\mathrm{Re}[\sigma(\omega')]}{\omega'-\omega}d\omega'~.
\end{equation}
%%%%%%%%
More specifically, as $\omega\rightarrow 0$, the imaginary part behaves as $Im(\sigma)\sim n_s/\omega$, and according to Kramers-Kronig relations, the real part has the form $Re(\sigma)\sim\pi n_s\delta(\omega)$. Here the coefficient $n_s$ of the delta function is defined as the superfluid density. By fitting data near the critical temperature, we find that with various couplings, the superfluid density has the behaviour
\begin{equation}\label{eq:ns}
n_s\simeq C_1 T_c(1-T/T_c)~,
\end{equation}
which means that $n_s$ vanishes linearly as $T$ goes to $T_c$. This is consistent with that happens in the minimal coupling. The various values of the coefficient $C_1$ are listed in Table {\ref{tableTc}}. We see that $C_1$ decreases drastically to suppress the superfluid density when the coupling is increased. This property is consistent with the condensation shown in Fig. \ref{condensate} that the stronger coupling corresponds to lower condensation gap.
%%%%%%%
\begin{table}
  \centering
  \begin{tabular}{|c||c|c|c|c|c|c|c|}
    \hline
    $\lambda$&0 & 0.2 &0.4 &  0.6&0.63 &0.633&0.6339 \\ \hline
    $C_1$&$16.92$ & $10.95$&$4.04$&$0.46$&$0.09$&$0.04$&$0.01$ \\ \hline
  \end{tabular}
  \caption{\label{tableTc}The coefficient $C_1$ of the superfluid density near the critical temperature for different coupling.}
\end{table}

\section{Conclusions and Discussions}
\label{sectCon}
In this work we carried out a holographic realization of the spontaneous scalarization in the Einstein-scalar-Gauss-Bonnet gravity theory with a negative cosmological constant. We first studied the stability of this theory with or without the presence of an electromagnetic field. Perturbing the background metric of a Schwarzschild-AdS or a Reissner-Nordstrom-AdS black hole with planar horizons by a probe scalar field, we calculated the effective mass. We found that this effective mass becomes negative for non-zero values of the charge of the scalar field and the coupling $\lambda$ of the scalar field to the GB term. We then studied  the holographic scalarization of the background black holes in the presence of neutral and charged scalar field, respectively.

In the case the scalar field is neutral, when the GB coupling $\lambda$ is tuned to be large enough, a hairy black hole could form and we numerically construct the hairy solution in the bulk theory. From gauge/gravity duality, the formation of hairy black hole in the bulk corresponds to the condensation becoming  nonzero, and this could be treated as certain holographic phase transition in the dual boundary theory even though it occurs without any breaking of symmetry. We then probe the phase transition by computing the $\lambda$ dependent vacuum expectation value of the dual scalar operator and the entanglement entropy in the boundary theory.

We know that a hairy black hole can be formed in AdS space if a charged scalar field couples to a Maxwell field, accompanying with spontaneous breaking of the $U(1)$ symmetry in Abelian Higgs model \cite{Gubser:2008px}. This process was described by a holographic superconducting phase transition \cite{Hartnoll:2008vx,Hartnoll:2008kx} in the boundary theory.  However, apparently, the formation of the hairy black hole because of GB coupling in this model is different from that with charged scalar. We have discussed the differences both from the bulk  side and on the boundary theory side. In the bulk side, the occurrence of scalarization in our model stems from that the GB coupling term could lower the effective mass of the scalar field and cause instability, while in Abelian Higgs model it is the non-vanishing $U(1)$ gauge field contributes a negative effect to the effective mass. In the boundary theory, in our model the tuning quantity is the coupling parameter and the VEV is nonzero above the critical value $\lambda_c$. Moreover, the phase transition should be a quantum type because we do not have any symmetry involved. However, the formation of a hairy black hole of Abelian Higgs model is dual to holographic superconductor phase transition which accompanies with the breaking of $U(1)$ gauge symmetry. The tuning quantity is the temperature of the theory, and above a critical temperature the system is dual to a normal state with vanishing VEV while it is dual to a holographic superconducting state.

Finally, we investigated the holographic scalarization of the charged scalar field and it was found that the background black holes are scalarized below a critical temperature. In the probe limit, the temperature dependent properties of the scalar condensation were studied  and the optical conductivity of superconducting phase transition was analyzed. It was found that  as the charge of the scalar field is increased the critical temperature also increases.
For the coupling of the scalar field to the GB term $\lambda$ it was found that there exist a critical value of $\lambda_c$ beyond which a scalarized black hole does not form in the gravity sector while on the boundary there is no any holographic superconducting phase. This can be understood from the fact that above $\lambda_c$ the gravitational attraction from the GB high curvature term becomes stronger and the formation of the scalarized black
hole is not possible. Also below $\lambda_c$ the conductivity and the superfluid density was calculated. We found that as $\lambda$ becomes stronger, reaching its critical value $\lambda_c$, the real part of conductivity increases to unit for all frequency, which means that the conductivity is weaker and finally the system is dual to a metal at fixed temperature.

To conclude, the (in)stability analysis in section \ref{secII} indicates that the combined effect of two different scalarization mechanisms (interaction between the scalar field and GB curvature correction and interaction between the scalar and the $U(1)$ electromagnetic field) accommodates a wider and deeper effective mass to speed up the formation of hairy black holes. Correspondingly, in the boundary theory, we shown that above certain critical temperature, only certain phase transition intrigued by large enough GB coupling could occur such that the scalar hair forms. However, when temperature drops below a critical value, the holographic superconducting condensation participates and we have the combined stronger effects on the formation of hairy black holes.

It would be interesting to investigate to what holographic system on the boundary the formation of a condensation corresponds without breaking a symmetry in the bulk. As we already discussed this may correspond to certain quantum  phase transition which occurs in the dual theory. It is known that the coupling of the scalar field to the GB term in flat spacetime can lead to a scalarized black hole solution violating in this way the non-hair theorems. Therefore, it  would also be interesting to investigate in the AdS spacetime the dynamics of the  interplay between the gravitational force displayed by the GB term and the electromagnetic field in the formation of a  scalarized black hole solution.

\section*{\bf Acknowledgements}

This work is supported by the Natural Science Foundation
of China under Grants No. 11705161, No. 11775036, and
No. 11847313, and Fok Ying Tung Education Foundation
under Grant No. 171006.  Jian-Pin Wu is also supported by Top
Talent Support Program from Yangzhou University.

%%%%%%%%%%%%%%%%%%%%%%%%%%%% BIBLIOGRAPHY %%%%%%%%%%%%%%%%%%%%%%%%%%%%%%%%%%%%%%

\end{document}